\documentclass[aps,prb,preprint,showpacs,preprintnumbers,amsmath,amssymb,floatfix]{revtex4-1} 

\usepackage[dvipdfmx]{graphicx}
\usepackage{bm}
\usepackage{slashbox,multirow}

\newcommand{\be}{\begin{equation}}
\newcommand{\ee}{\end{equation}}
\newcommand{\eq}[1]{Eq.~(\ref{#1})}
\newcommand{\fig}[1]{Fig.~\ref{#1}}
\def\bea{\begin{eqnarray}}
\def\eea{\end{eqnarray}}

\def\bra{\langle}
\def\ket{\rangle}
\def\vq{{\bf q}}

\def\vk{{\bf k}}
\def\vQ{{\bf Q}}

\begin{document}

\title{Electron self-energy from quantum charge fluctuations in the layered {\boldmath $t$}-{\boldmath $J$} model 
with long-range Coulomb interaction} 

\author{Hiroyuki Yamase$^{1,2}$, Mat\'{\i}as Bejas$^{3}$, and Andr\'es Greco$^{3}$}
\affiliation{
{$^1$}International Center of Materials Nanoarchitectonics, 
National Institute for Materials Science, Tsukuba 305-0047, Japan\\
{$^2$}Department of Condensed Matter Physics, Graduate School of Science, Hokkaido University, Sapporo 060-0810, Japan\\
{$^3$}Facultad de Ciencias Exactas, Ingenier\'{\i}a y Agrimensura and
Instituto de F\'{\i}sica Rosario (UNR-CONICET),
Av. Pellegrini 250, 2000 Rosario, Argentina
}

\date{\today}

\begin{abstract}
Employing a large-$N$ scheme of the layered $t$-$J$ model with the long-range Coulomb interaction, which 
captures fine details of the charge excitation spectra recently observed in cuprate superconductors, 
we explore the role of charge fluctuations on the electron self-energy. 
We fix temperature at zero and focus on quantum charge fluctuations. 
We find a pronounced asymmetry of the imaginary part of the 
self-energy Im$\Sigma(\vk, \omega)$ with respect to $\omega=0$, which is driven by strong 
electron correlation effects. 
The quasiparticle weight is reduced dramatically, which occurs almost isotropically along 
the Fermi surface.  
Concomitantly an incoherent band and a sharp side band are generated 
and acquire sizable spectral weight. 
All these features are driven by usual on-site charge fluctuations, which 
are realized in a rather high-energy region and yield plasmon excitations. 
On the other hand, the low-energy region with the scale 
of the superexchange interaction $J$ is dominated by bond-charge fluctuations. 
Surprisingly, compared with the effect of on-site charge fluctuations, 
their effect on the electron self-energy is much weaker, even if the system 
approaches close to bond-charge instabilities. 
Furthermore, quantum charge dynamics does not produce a clear kink nor a pseudogap 
in the electron dispersion. 
\end{abstract}

\maketitle
\section{introduction}
High-temperature cuprate superconductors are realized upon charge-carrier doping 
into the antiferromagnetic Mott insulators. 
Many studies highlighted the magnetic properties. In particular, 
the effect of spin fluctuations has been the major subject to explore the cuprate phenomenology 
such as high-temperature superconductivity, the anomalous metallic state including 
the enigmatic pseudogap, and others \cite{keimer15}. 
Needless to say, the charge dynamics has also been widely recognized as an important subject. 
However, much was not known about the charge dynamics in momentum 
and energy space. Recently, advanced x-ray scattering techniques \cite{ishii05,ghiringhelli12,chang12,achkar12,da-silva-neto15,wslee14,ishii14} 
changed this situation. 

For electron-doped cuprates (e-cuprates), charge excitations were reported 
in both low-energy region \cite{da-silva-neto15,da-silva-neto16} with a scale 
of the superexchange interaction $J (\approx$ 150 meV) and 
high-energy region \cite{ishii05,wslee14,ishii14,dellea17,hepting18,jlin20} with a scale larger than $J$. 
These features were captured in the layered $t$-$J$ model with the long-range Coulomb interaction 
as a dual structure of the charge excitations \cite{bejas17}. 
The low-energy charge excitations originate mainly from the bond-charge fluctuations \cite{yamase15b,yamase19} 
and the high-energy ones from acousticlike plasmons \cite{greco16,greco19,greco20}. 

For hole-doped cuprates (h-cuprates), mainly low-energy charge fluctuations were 
reported \cite{ghiringhelli12,chang12,achkar12} and their origin is under debate despite intensive theoretical 
studies \cite{bejas12,allais14,meier14,wang14,atkinson15,yamakawa15,mishra15,zeyher18}. 
The origin of high-energy charge excitations in h-cuprates \cite{ishii17} is also highly controversial. 
High-energy excitations can be i) specific to e-cuprates \cite{wslee14,dellea17}, 
ii) present as a broad peak of the particle-hole excitation spectrum \cite{ishii17}, 
iii) present as plasmons similar to the e-cuprates \cite{greco19,hepting18,jlin20,singh20,nag20}, and iv) 
related to the strange metal physics \cite{mitrano19,husain19}, not to plasmons.  

Cuprate superconductors are strongly correlated electron systems and the bare hopping integral 
$t$, whose energy scale is usually eV, is renormalized to be in the scale of $J$, leading to a 
relatively narrow band width. In this case, a coupling to bosonic excitations 
such as charge fluctuations revealed recently is expected to generate anomalous features 
in the electron self-energy. 

However, there are only a few studies along that motivation. 
Among others, Ref.~\onlinecite{markiewicz07a} studied the effect of the optical plasmon on the electron dispersion 
in a phenomenological framework. 
Reference~\onlinecite{zemljic08} discussed the effect of incoherent hole motion to explain 
the so-called high-energy kink of the electron 
dispersion observed by angle-resolved photoemission spectroscopy (ARPES) around $-(0.3 \sim 1)$ eV 
(Ref.~\onlinecite{graf07,xie07,valla07,meevasana07,chang07,zhang08}). 
We still only have sparse insights into the effect of charge fluctuations on the one-particle properties. 
Hence it is very interesting to apply the large-$N$ theory, which captures fine details 
of charge excitation spectra observed by experiments, to the study of the electron self-energy 
and to perform a comprehensive analysis of 
the one-particle properties of electrons in a microscopic model. 
Such a study will potentially open a path 
to study the charge dynamics in cuprates also from 
the one-particle properties, which may accelerate our understanding of the cuprate physics.

In this paper, we employ the 
layered $t$-$J$ model with the long-range Coulomb interaction in a large-$N$ scheme 
and study the electron self-energy from charge fluctuations. 
This is a challenging study because we have to go beyond leading order in a large-$N$ expansion. 
We fix temperature at zero and focus on quantum charge fluctuations. 
We shall clarify how the quantum charge dynamics renormalizes the quasiparticle dispersion 
and generates bands for both e- and h-cuprates, including a possibility of kinks and a pseudogap 
in the electron dispersion. 
Since the charge dynamics consists of on-site charge and bond-charge fluctuations, 
we sharply distinguish these two features in the present analysis. 

The present paper is organized as follows. 
After providing a brief summary of the formalism in Sec. II, 
we present results in Sec. III. Starting from a summary of the typical 
charge excitation spectrum,  we show the self-energy effect 
from on-site charge fluctuations, which describe plasmons, 
and then clarify the effect of bond-charge fluctuations. 
The obtained results are discussed from a view of both theoretical 
and experimental perspectives in Sec. IV. Concluding remarks are given in Sec. V. 
Appendices consist of two parts: a full description of the formalism 
and additional results for e-cuprates.

\section{Model and Formalism} 
As a microscopic model of cuprate superconductors, 
we study the $t$-$J$ model on a square lattice by including interlayer hopping and 
the long-range Coulomb interaction. The Hamiltonian is given by 
\begin{equation}
H = -\sum_{i, j,\sigma} t_{i j}\tilde{c}^\dag_{i\sigma}\tilde{c}_{j\sigma} + 
\sum_{\langle i,j \rangle} J_{ij} \left( \vec{S}_i \cdot \vec{S}_j - \frac{1}{4} n_i n_j \right)
+\frac{1}{2} \sum_{i,j} V_{ij} n_i n_j \,,
\label{tJV}  
\end{equation}
where $\tilde{c}^\dag_{i\sigma}$ ($\tilde{c}_{i\sigma}$) are 
the creation (annihilation) operators of electrons with spin $\sigma (=\uparrow, \downarrow)$  
in the Fock space without double occupancy at any site,  
$n_i=\sum_{\sigma} \tilde{c}^\dag_{i\sigma}\tilde{c}_{i\sigma}$ 
is the electron density operator, $\vec{S}_i$ is the spin operator, and 
the sites $i$ and $j$ run over a three-dimensional lattice. 
The hopping $t_{i j}$ takes the value $t$ $(t')$ between the first- (second-) nearest-neighbor 
sites on the square lattice and it is scaled by $t_z$ between the layers; see Eqs.~(\ref{Epara}) and (\ref{Eperp}) for the
explicit form of the electron dispersion.
$\langle i,j \rangle$ denotes the nearest-neighbor sites and 
the exchange interaction $J_{i j}=J$ is considered only inside the plane. 
We neglect the exchange term between the planes, which is much smaller than $J$
(Ref.~\onlinecite{thio88}).
$V_{ij}$ is the long-range Coulomb interaction on the three-dimensional lattice 
and its functional form is given in momentum space later [\eq{LRC}]. 
While cuprates are essentially two-dimensional systems, it is crucial to employ 
the three-dimensional model, namely the layered model, because the long-range Coulomb interaction leads 
to a sizable momentum dependence of plasmons along the $q_z$ direction as was shown 
in classical papers \cite{grecu73,fetter74,grecu75}. 

It is highly nontrivial to treat the $t$-$J$ model (\ref{tJV}) because of the local constraint 
prohibiting the double occupancy at any site. 
Here we employ a large-$N$ technique in a path integral representation in terms of the Hubbard 
operators \cite{foussats02}.  
In the large-$N$ scheme, the number of spin components is extended from 2 to $N$ 
and physical quantities are computed by counting the power of $1/N$ systematically. 
One of the advantages of this method is that it treats all 
possible charge excitations on an equal footing \cite{bejas12,bejas14}, which 
makes this method potentially 
interesting in light of the new x-rays experiments revealing excitation spectra 
in the pure charge channel \cite{ishii05,ghiringhelli12,chang12,achkar12,wslee14,ishii14, da-silva-neto15,da-silva-neto16,dellea17,ishii17,hepting18,jlin20,singh20,nag20}.

The computation of the electron self-energy requires extended calculations in a large-$N$ framework 
beyond the leading order theory \cite{wang91,merino03,bejas06}. 
Since we would like to highlight the effect of 
charge fluctuations on the self-energy in this paper, 
we present the most essential part of the formalism here, 
leaving its full description to Appendix~A. 
In short, we first obtain the electron dispersion at leading order 
and compute charge fluctuations at order of $1/N$. 
We then calculate their contributions to the electron self-energy at 
the same order.

At leading order, the electron dispersion $\varepsilon_{\vk}$ is obtained as 
\be
\varepsilon_{\vk} = \varepsilon_{\vk}^{\parallel}  + \varepsilon_{\vk}^{\perp} \,,
\label{xik}
\ee
where the in-plane dispersion $\varepsilon_{\vk}^{\parallel}$ and the out-of-plane dispersion 
$\varepsilon_{\vk}^{\perp}$ are given, respectively, by
\be
\varepsilon_{\vk}^{\parallel} = -2 \left( t \frac{\delta}{2}+\Delta \right) (\cos k_{x}+\cos k_{y})-
4t' \frac{\delta}{2} \cos k_{x} \cos k_{y} - \mu \,,\\
\label{Epara}
\ee
\be
\varepsilon_{\vk}^{\perp} = - 2 t_{z} \frac{\delta}{2} (\cos k_x-\cos k_y)^2 \cos k_{z}  \,. 
\label{Eperp}
\ee
Here we measure the in-plane momenta $k_x$ and $k_y$ and the out-of-plane momentum $k_z$ 
in units of $a^{-1}$ and $d^{-1}$, respectively; 
$a$ ($d$) is the  lattice constants in the plane (distance between the planes). 
While the dispersions are similar to noninteracting ones, the hopping integrals $t$, $t'$, and $t_z$ 
are renormalized by a factor of $\delta/2$ where $\delta$ is a doping rate. 
The quantity $\Delta$ in \eq{Epara} is a mean-field value of the bond field and is given by 
\bea{\label {Delta}}
\Delta = \frac{J}{4N_s N_z} \sum_{\vk} (\cos k_x + \cos k_y) n_F(\varepsilon_\vk) \; , 
\eea
where $n_F(\varepsilon_\vk)$ is the Fermi function, $N_s$ the total 
number of lattice sites on the square lattice, and $N_z$ the number of layers 
along the $z$ direction. For a given doping $\delta$, 
$\mu$ and $\Delta$ are determined self-consistently by solving \eq{Delta} and 
\be
(1-\delta)=\frac{2}{N_s N_z} \sum_{\vk} n_F(\varepsilon_\vk)\,.
\ee

Charge fluctuations included in the $t$-$J$ model (\ref{tJV}) are described by the 
$6 \times 6$ matrix of the bosonic propagator at the order of $1/N$: 
\be
[D_{ab}(\vq,\mathrm{i}\nu_n)]^{-1} 
= [D^{(0)}_{ab}(\vq,\mathrm{i}\nu_n)]^{-1} - \Pi_{ab}(\vq,\mathrm{i}\nu_n)\,,
\label{dyson}
\ee
where $a$ and $b$ run from 1 to 6; $\vq$ is a three-dimensional wavevector and 
$\nu_n$ is a bosonic Matsubara frequency. 
$D^{(0)}_{ab}(\vq,\mathrm{i}\nu_n)$ is the bare bosonic propagator and is obtained as 
\begin{widetext}
\begin{eqnarray}
[D^{(0)}_{ab}({\bf q},\mathrm{i}\nu_{n})]^{-1}= N \left(
 \begin{array}{cccccc}
\frac{\delta^2}{2} \left[ V(\vq)-J(\vq)\right]
& \frac{\delta}{2} & 0 & 0 & 0 & 0 \\
   \frac{\delta}{2}  & 0 & 0 & 0 & 0 & 0 \\
   0 & 0 & \frac{4}{J}\Delta^{2} & 0 & 0 & 0 \\
   0 & 0 & 0 & \frac{4}{J}\Delta^{2} & 0 & 0 \\
   0 & 0 & 0 & 0 & \frac{4}{J}\Delta^{2} & 0 \\
   0 & 0 & 0 & 0 & 0 & \frac{4}{J}\Delta^{2} \,
 \end{array}
\right).
\label{D0}
\end{eqnarray}
\end{widetext}
Here $J(\vq) = \frac{J}{2} (\cos q_x +  \cos q_y)$  and 
$V(\vq)$ is the long-range Coulomb interaction in momentum space for a layered system, 
\be
V(\vq)=\frac{V_c}{A(q_x,q_y) - \cos q_z} \,,
\label{LRC}
\ee
where $V_c= e^2 d(2 \epsilon_{\perp} a^2)^{-1}$ and 
\be
A(q_x,q_y)=\alpha (2 - \cos q_x - \cos q_y)+1 \,.
\label{Aq}
\ee
The Coulomb interaction $V(\vq)$ 
is easily obtained by solving Poisson's equation on the lattice \cite{becca96}. 
In \eq{Aq}, $\alpha=\frac{\tilde{\epsilon}}{(a/d)^2}$ and $\tilde{\epsilon}=\epsilon_\parallel/\epsilon_\perp$; 
$\epsilon_\parallel$ and $\epsilon_\perp$ are the 
dielectric constants parallel and perpendicular to the planes, respectively;
$e$ is the electric charge of electrons.

The $6 \times 6$ matrix $\Pi_{ab}$ is the bosonic self-energy
\begin{eqnarray}
&& \Pi_{ab}(\vq,\mathrm{i}\nu_n)
            = -\frac{N}{N_s N_z}\sum_{\vk} h_a(\vk,\vq,\varepsilon_\vk-\varepsilon_{\vk-\vq}) 
            \frac{n_F(\varepsilon_{\vk-\vq})-n_F(\varepsilon_\vk)}
                                  {\mathrm{i}\nu_n-\varepsilon_\vk+\varepsilon_{\vk-\vq}} 
            h_b(\vk,\vq,\varepsilon_\vk-\varepsilon_{\vk-\vq}) \nonumber \\
&& \hspace{25mm} - \delta_{a\,1} \delta_{b\,1} \frac{N}{N_s N_z}
                                       \sum_\vk \frac{\varepsilon_\vk-\varepsilon_{\vk-\vq}}{2}n_F(\varepsilon_\vk) \; , 
\label{Pi}
\end{eqnarray}
where the six-components vertex is given by  
\begin{align}
 h_a(\vk,\vq,\nu) =& \left\{
                   \frac{2\varepsilon_{\vk-\vq}+\nu+2\mu}{2}+
                   2\Delta \left[ \cos\left(k_x-\frac{q_x}{2}\right)\cos\left(\frac{q_x}{2}\right) +
                                  \cos\left(k_y-\frac{q_y}{2}\right)\cos\left(\frac{q_y}{2}\right) \right];
                                                   \right. \nonumber \\
               & \hspace{-10mm} \left. 1; -2\Delta \cos\left(k_x-\frac{q_x}{2}\right); -2\Delta \cos\left(k_y-\frac{q_y}{2}\right);
                         2\Delta \sin\left(k_x-\frac{q_x}{2}\right);  2\Delta \sin\left(k_y-\frac{q_y}{2}\right)
                 \right\} \, .
\label{vertex-h}
\end{align}
Here the dependence on $k_z$ and $q_z$ enters only through $\epsilon_{\vk-\vq}$ in the first column 
and the other columns contain the in-plane momentum only. 

Charge fluctuation spectra are obtained by the analytical continuation in \eq{dyson}
\be
\mathrm{i}\nu_n \rightarrow \nu + \mathrm{i} \Gamma_{\rm ch}\,,
\label{gamma-ch}
\ee
where $\Gamma_{\rm ch} (>0)$ is infinitesimally small. 
By studying Im$D_{ab}(\vq, \nu)$ we can elucidate all possible 
charge dynamics in the layered $t$-$J$ model as was already performed in 
the literature \cite{bejas17}.

If we set $J=0$, we would obtain $\Delta=0$ and all fluctuations associated with the bond field vanish. 
The bosonic propagator $D_{ab}$ is reduced to a $2\times2$ matrix with $a,b=1,2$ 
and only usual on-site charge fluctuations are active [see also \eq{deltaXa}]. 
In fact, the element $(1,1)$ of $D_{ab}$ is related to 
the usual charge-charge correlation function \cite{foussats02}.  
$D_{22}$ and $D_{12}$ correspond to fluctuations associated with the non-double-occupancy condition 
and correlations between non-double-occupancy condition and charge density fluctuations, respectively. 
When $J$ is finite,  bond-charge fluctuations become active and 
$a$ and $b$ take values from $1$ to $6$. Thus, both on-site charge and bond-charge fluctuations 
are present for a realistic situation.

Although each lattice site in the $t$-$J$ model corresponds to 
a Cu atom in the CuO$_2$ plane, the effect of O atoms is implicitly included because 
the $t$-$J$ model is derived from the three-band Hubbard model in the strong 
coupling limit \cite{fczhang88}. 
Thus, naively speaking, the on-site charge fluctuations in the $t$-$J$ model, which is 
described by the $2\times 2$ sector of $D_{ab}$, 
correspond to charge fluctuations  
on each Cu site, whereas  
the bond-charge fluctuations described by $D_{ab}$ with $a,b=3$-$6$ 
account for charge fluctuations between Cu sites, i.e., at the O sites. 
However, this does not mean that the on-site charge and bond-charge 
fluctuations can be considered only around Cu sites and O sites, respectively, because 
Cu orbitals strongly hybridize with O orbitals, forming the Zhang-Rice singlet \cite{fczhang88}. 

Next, we compute the electron self-energy. 
At the order of $1/N$, the imaginary part of the self-energy is obtained in a compact form as \cite{bejas06} 
\begin{equation}
{\mathrm{Im}}\Sigma({\mathbf{k}},\omega) = \frac{-1}{N_{s} N_z}
\sum_{{\mathbf{q}}} \sum_{a,b} {\rm Im}D_{ab} (\vq,\nu) h_{a}(\vk,\vq,\nu) 
h_{b}(\vk,\vq,\nu) \left[
n_{F}( -\varepsilon_{\vk-\vq}) +n_{B}(\nu) 
\right] \, ,
\label{ImSig}
\end{equation}
where $\nu=\omega - \varepsilon_{\vk-\vq}$ and $n_B(\nu)$ is the Bose function. 
Note that the self-energy effects arise from charge fluctuations described by $D_{ab}$ 
and the effect of spin fluctuations does not appear at $O(1/N)$. 
The form of Im$\Sigma(\vk,\omega)$ in \eq{ImSig} has the same structure as a self-energy 
obtained from the Fock diagram in a usual perturbation theory. 
However, in the large-$N$ scheme, we have both Hartree and Fock diagrams in a nontrivial 
way at order of $1/N$; see Appendix~A for further details.

Using the Kramers-Kronig relations, we obtain $\mathrm{Re} \Sigma(\vk,\omega)$ 
from $\mathrm{Im}\Sigma(\vk,\omega)$. 
Since the electron Green's function $G(\vk,\omega)$ is given by 
\be
G^{-1}(\vk,\omega) = \omega + \mathrm{i} \Gamma_{\rm sf} -\varepsilon_{\vk} - \Sigma (\vk,\omega) \,,
\ee
we can compute the one-particle spectral function  
$A({\bf k},\omega)=-\frac{1}{\pi} {\rm Im}G({\bf k},\omega)$ as 
\be
A({\bf k},\omega)= -\frac{1}{\pi} \frac{{\rm Im}\Sigma({\bf k},\omega) - \Gamma_{\rm sf}}
{[\omega- \varepsilon_{\vk}-{\rm Re}\Sigma({\bf k},\omega)]^2 
+ [-{\rm Im}\Sigma(\vk,\omega) +\Gamma_{\rm sf}]^2} \,,
\label{Akw}
\ee
where $\Gamma_{\rm sf}(>0)$ originates from the analytical continuation 
in the electron Green's function. 

The real part $\mathrm{Re} \Sigma(\vk,\omega)$ contains a frequency-independent contribution and 
shifts the band dispersion.  Since the dispersion in \eq{xik} is chosen to reproduce the experimental Fermi surface, 
we drop this frequency-independent contribution to avoid double counting. 
This procedure does not affect anything on $\mathrm{Im}\Sigma(\vk,\omega)$, which can be 
easily understood from the Kramers-Kronig relations.

We have formulated a theory in the large-$N$ scheme to compute the electron self-energy 
at the order of $1/N$. This kind of calculations is not seen much in literature. For example, 
the large-$N$ theory is often formulated in the slave-boson method.  
However there are few calculations including fluctuations above the
mean field \cite{wang91,wang92}, and many calculations were performed 
at the mean-field level \cite{kotliar88,morse91,wang92}.

\section{Results}
Following our previous works, which capture  
the charge dynamics in e-cuprates in both high- \cite{greco16,bejas17,greco19,greco20,nag20} 
and low-energy region \cite{yamase15b,bejas17,yamase19} consistently, 
we choose the band parameters as 
$t'/t=0.3$, $t_z/t=0.1$.  The other parameters are taken to be  
$J/t=0.3$, $V_c/t=17$, $\alpha=4.5$, and $N_z=10$. 
The resulting momentum along the $z$ direction is given by 
a discrete value such as $q_z=2\pi n_z/10$ with $n_z$ being integer. 
While a value of $\Gamma_{\rm ch}$ is infinitesimally small in \eq{gamma-ch}, 
we set  $\Gamma_{\rm ch}/t=0.03$ for numerical reasons. 
This finite $\Gamma_{\rm ch}$ may mimic phenomenologically a broadening of the spectrum \cite{greco19} 
due to the resolution of experimental measurements and also additional effects not 
included in the present theory such as higher-order electron correlation effects
\cite{prelovsek99} and coupling to other degrees of freedom. 
Temperature is set at $T=0$ and we focus on quantum charge fluctuations. 
We first present results for the doping rate $\delta=0.15$, and  
then study the doping dependence later. 
When computing the spectral function, 
we choose $\Gamma_{\rm sf}/t=0.01$ to make the spectrum reasonably broadened.

Complete information of the charge dynamics is included in the $6 \times 6$ matrix $D_{ab}$ 
in Eq.~(\ref{dyson}),  
where both on-site charge and bond-charge fluctuations are present. 
As we shall show below, the major contributions to the electron self-energy 
come from the on-site charge fluctuations described by the $2 \times 2$ sector of $D_{ab}$,  
namely $a,b=1,2$ in \eq{ImSig}.
Hence we first focus on results from on-site charge fluctuations and clarify 
the major features of the electron self-energy due to the coupling to them. 
The effect of bond-charge fluctuations and a possible pseudogap feature are studied afterward. 
We also study a different set of band parameters appropriate to h-cuprates. 
We mention some care about our results for e-cuprates when we compare them 
with experiments. 
Finally we present results in the absence of the long-range Coulomb interaction 
to highlight its role in the electron self-energy.

\subsection{Charge excitation spectrum} 
The charge excitation spectra Im$D_{ab}(\vq,\nu)$ are already presented 
for all components $a$ and $b$ in Figs. 1 and 8 in Ref.~\onlinecite{bejas17}. 
Furthermore, detailed analyses were performed 
for both the on-site charge excitations \cite{greco16,bejas17,greco19,greco20} and 
the bond-charge excitations \cite{yamase15b,bejas17,yamase19}. 
For the sake of readers' convenience, we here summarize the typical features of 
charge excitation spectra relevant to the understanding of the electron self-energy. 

\begin{figure}[ht]
\centering
\includegraphics[width=9cm]{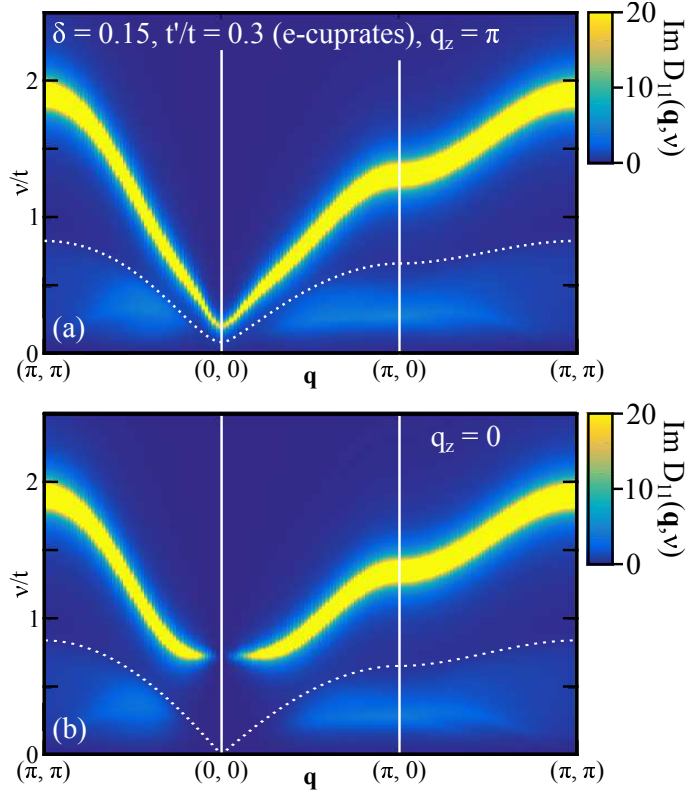}
\caption{(Color online) 
Typical charge excitation spectrum [intensity map of Im$D_{11}(\vq,\nu)$]  
in the plane of 
momentum $\vq$ and energy $\nu$ for $q_z=\pi$ (a) and $q_z=0$ (b). 
The result for $q_z=\pi$ is a representative one of results for other $q_z (\ne 0)$.  
The dotted curve is the upper limit of the continuum excitations. 
}
\label{ImD11}
\end{figure}

The diagonal components of Im$D_{ab}(\vq,\nu)$ are always positive as they should be. 
Figure~\ref{ImD11} shows a typical charge-excitation spectrum by choosing $a=b=1$. 
The sharp features correspond to plasmon excitations. 
In particular, they exhibit a V-shaped dispersion around $\vq=(0,0,q_z)$ with $q_z \ne 0$ 
[\fig{ImD11}(a)].   
The gap at $\vq=(0,0,q_z)$ is proportional to the interlayer hopping integral $t_z$ (Refs.~\onlinecite{grecu73,grecu75,greco16}).  
For $q_z=0$ [\fig{ImD11}(b)], instead, 
we obtain the well-known optical plasmon with a nearly flat dispersion 
around $\omega =0$.  In the low-energy region, namely, below the dotted line in \fig{ImD11}, 
the so-called particle-hole continuum is realized. 
No strong spectral weight appears inside the continuum 
and there is no tendency toward conventional charge-density-wave order including stripes. 
When we choose $a=b=3$-$6$ and explore bond-charge excitations, we find 
strong spectral weight in the low-energy region inside the continuum \cite{bejas17}.  
However, the effect of the bond-charge excitations on the electron self-energy 
is minor as we shall show below. 

The off-diagonal components Im$D_{ab}(\vq,\nu)$ are not positive definite any longer 
(see Figs. 1 and 8 in Ref.~\onlinecite{bejas17}). 
Typically they exhibit strong spectral weight in almost the same $(\vq,\nu)$ region 
where Im$D_{11}(\vq,\nu)$ exhibit the plasmon dispersion.

\subsection{Self-energy effect from on-site charge fluctuations} 
\begin{figure}[b]
\centering
\includegraphics[width=9cm]{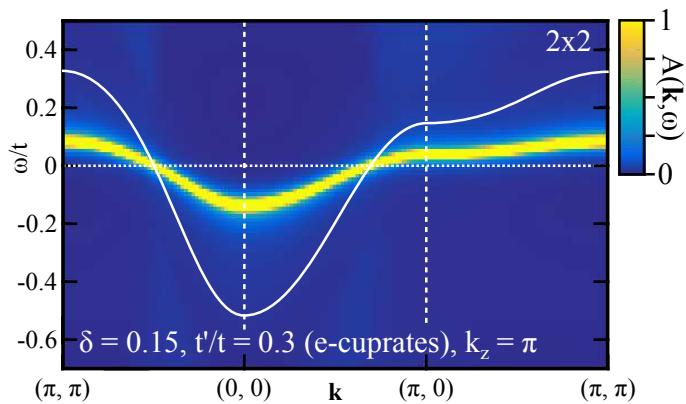}
\caption{(Color online) 
Intensity map of $A(\vk,\omega)$ along the direction 
$(\pi,\pi)$-$(0,0)$-$(\pi,0)$-$(\pi,\pi)$ for $k_z=\pi$. 
The quasiparticle dispersion in the presence of the self-energy is in yellow. 
The white curve is the quasiparticle dispersion 
obtained in leading order theory without the self-energy [\eq{xik}]. 
See also \fig{Akw-ph}(a) obtained after a particle-hole transformation. 
}
\label{QPdispersion}
\end{figure}

The quasiparticle dispersion in leading order theory [\eq{xik}] is plotted with a white curve in \fig{QPdispersion}, 
where the hopping integrals are already renormalized and the energy scale is reduced to $J(=0.3t)$. 
The quasiparticle dispersion is further renormalized by the coupling to the on-site charge fluctuations. 
To see that, we show in \fig{QPdispersion} 
the intensity map of $A(\vk,\omega)$ [\eq{Akw}] around the Fermi energy for $k_z=\pi$; 
the $k_z$ dependence is weak and essentially the same results are obtained for a different value of $k_z$. 
The band width is suppressed significantly to become less than half, 
indicating that the quasiparticle weight $Z$ is reduced substantially by the on-site charge fluctuations.

\begin{figure}[ht]
\centering
\includegraphics[width=12cm]{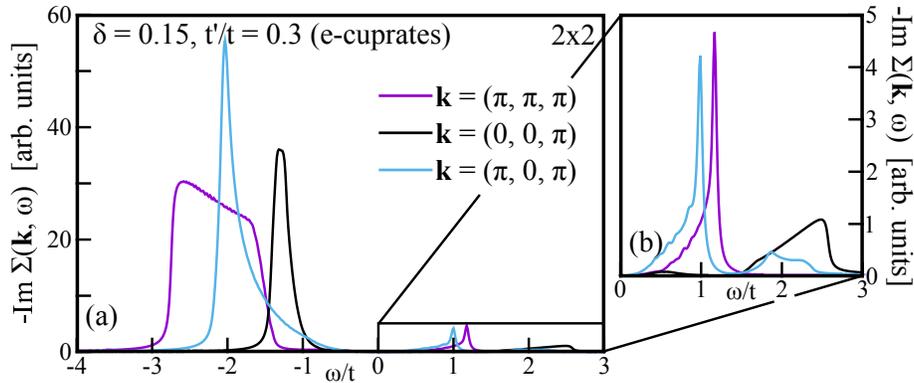}
\caption{(Color online) 
Imaginary part of the electron self-energy, $-{\rm Im}\Sigma(\vk,\omega)$, as a function of $\omega$ 
for several choices of $\vk$. The positive energy region is magnified in the right panel. 
Results do not depend much on a value of $k_z$ and $k_z=\pi$  is taken as a representative one. 
See also \fig{ImSig-asym-ph} obtained after a particle-hole transformation. 
}
\label{ImSig-asym}
\end{figure}

Figure~\ref{ImSig-asym} shows the imaginary part of the electron self-energy Im$\Sigma(\vk,\omega)$ 
as a function of energy $\omega$ for several choices of momenta $\vk$. 
The self-energy exhibits a pronounced asymmetry with respect to $\omega=0$ and 
it is strongly suppressed in $\omega > 0$, leading to substantial breaking 
of the particle-hole symmetry. This asymmetry does not come from  the band structure effect 
due to the presence of $t'$, but from the strong correlation effect due to the local constraint inherent in the $t$-$J$ model. 
In fact, calculations in the random phase approximation (RPA) predict that Im$\Sigma (\vk, \omega)$  
is rather symmetric with respect to $\omega=0$ (Ref.~\onlinecite{foussats08}). 
Hence the analysis in RPA would lead to insights different from the present paper. 

To understand the reason of the particle-hole asymmetry in \fig{ImSig-asym}, 
we go back to \eq{ImSig}. 
First as we clarified in Ref.~\onlinecite{bejas17} and have also shown in \fig{ImD11}, 
the charge excitation spectrum  
Im$D_{ab}(\vq,\nu)$ with $a,b=1,2$ has a sizable spectral weight in $\nu/t \lesssim 1$ 
around $\vq=(0,0,q_z)$ and in a higher energy region typically around $\vq=(\pi,\pi,q_z)$; 
the $q_z$ dependence is minor. Im$D_{ab}(\vq,\nu)$ is an odd function with respect to $\nu$. 
For $\nu >0$, we have Im$D_{aa}(\vq,\nu) \geq 0$ for the diagonal components    
and Im$D_{ab}(\vq,\nu) \leq 0$ for the off-diagonal components 
(see Figs.~1 and 8 in Ref.~\onlinecite{bejas17}). 
The vertex function is given by $h_{1}(\vk,\vq,\nu) \propto \nu$ for a large $\nu$ and 
$h_{2}(\vk,\vq,\nu)=1$; see \eq{vertex-h}.  
We consider three different values of $\vk$ as typical ones: $\vk=(0,0,k_z), (\pi,\pi,k_z)$, and $(\pi,0,k_z)$. 
The term of $n_{F}( -\varepsilon_{\vk-\vq}) +n_{B}(\nu)$ 
takes a constant value depending on $\vk$, $\vq$, and $\nu$ as summarized in 
Table~\ref{selection-rule}, which works as a selection rule. 
Note that for $\nu<0$, the term $n_{F}( -\varepsilon_{\vk-\vq}) +n_{B}(\nu)$ 
can become negative, but the sign of Im$D_{ab}(\vq,\nu)$ also changes; 
consequently their product does not change the sign. 
Considering all these features, we recognize that all components $a,b=1,2$ contribute additively 
to the summation for $\nu < 0$ in \eq{ImSig} whereas the contribution from the diagonal 
components $a=b$ is largely cancelled by the off-diagonal components $a\ne b$ for $\nu > 0$, 
leading to the strong asymmetry of the self-energy as shown in \fig{ImSig-asym}. The off-diagonal components 
of Im$D_{ab}(\vq,\nu)$ originate from the strong correlation effect related to 
the Lagrange multiplier describing the nondouble occupancy of electrons at any site on the lattice. 

\begin{table}[tb]
\begin{center}
\begin{tabular}{c||c|c|c|c} 
\hline
\multirow{2}*{\backslashbox {$\vk$} {$\vq$}} 
 & \multicolumn{2}{c|} {$\sim (0,0,q_z)$} &  \multicolumn{2}{c} {$\sim (\pi,\pi,q_z)$}  \\
\cline{2-5}
 & {$\nu <0$\;} & {$\nu > 0$\; } & {$\nu<0$\;} & {$\nu>0$\;} \\
\hline \hline
$ \sim (0,0,k_z)$ & -1 &  0 & 0 & 1  \\ \hline
$ \sim (\pi,\pi,k_z)$ & 0 &  1 & -1 & 0 \\ \hline
$ \sim (\pi,0,k_z)$ & -1 or 0 & 0 or 1& -1 or 0&  0 or 1 \\ \hline
\end{tabular} 
\end{center}
\caption{Typical values of $n_{F}( -\varepsilon_{\vk-\vq}) +n_{B}(\nu)$ at $T=0$. 
They may depend sensitively on the precise values of $\vk$ and $\vq$ 
for $\vk \sim (\pi, 0, k_z)$,   
because the Fermi surface is located rather close to $\vk \approx (\pi,0,k_z)$. 
}
\label{selection-rule}
\end{table}

Recalling that we have the relation $\nu=\omega-\varepsilon_{\vk-\vq}$ in \eq{ImSig} and the value of 
$| \varepsilon_{\vk-\vq} |$ is less than $1t$ whereas the typical energy range that 
we are interested in \fig{ImSig-asym} is more than $1t$, 
the sign of $\nu$ becomes the same as that of $\omega$ independent of $\vk$ and $\vq$ in most of cases. 
Hence by associating the sign of $\omega$ with that of $\nu$, 
the selection rule in Table~\ref{selection-rule} 
serves to identify the origin of the structure of Im$\Sigma(\vk,\omega)$. 
For $\omega > 0$, the peak around $\omega/t \approx 1$ for $\vk=(\pi,\pi,k_z)$ 
comes from the plasmon excitations around $\vq \sim (0,0,q_z)$.  
A long tail on a lower energy side of the peak 
originates from the acousticlike plasmon mode. 
The structure around $\omega/t \approx 2.5$ for $\vk=(0,0,k_z)$ comes mainly from charge excitations 
around $\vq \sim (\pi,\pi,q_z)$. For an intermediate momentum such as $\vk=(\pi,0,q_z)$, 
Im$\Sigma(\vk,\omega)$ exhibits typically two structures around $\omega/t \approx 1$ and $2$; 
the former stems from plasmons and the latter from charge excitations around 
$\vq \sim (\pi,\pi, q_z)$. 
For $\omega <0$, on the other hand, the major contribution of charge fluctuations becomes 
{\it vice versa}: charge fluctuations around $\vq \sim (0,0,q_z)$ and $(\pi,\pi,q_z)$ 
form the structure around $\omega/t \approx - 1$ for $\vk=(0,0,k_z)$ and 
$\omega/t \approx - 2$ for $\vk=(\pi,\pi,k_z)$, 
respectively. For an intermediate momentum $\vk=(\pi,0,k_z)$, essentially a single peak 
is realized around $\omega/t \approx -2$ 
with a sizable tail on the side of $\omega=0$.

\begin{figure}[ht]
\centering
\includegraphics[width=8cm]{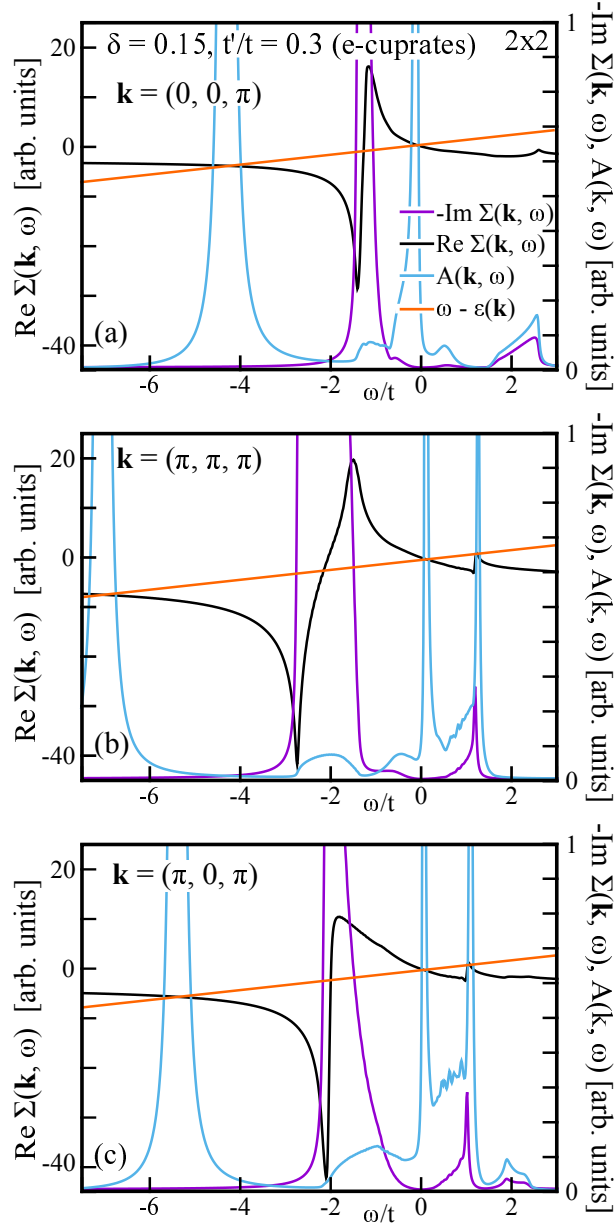}
\caption{(Color online) 
Energy dependence of $- {\rm Im}\Sigma(\vk,\omega)$, ${\rm Re}\Sigma(\vk,\omega)$, 
and $A(\vk,\omega)$ for $\vk=(0,0,\pi)$ (a), $(\pi,\pi,\pi)$ (b), and  $(\pi, 0,\pi)$ (c). 
The line of $\omega - \varepsilon (\vk)$ is also drawn. 
The scales of $- {\rm Im}\Sigma(\vk,\omega)$ and $A(\vk, \omega)$ 
correspond to the right vertical axis and their units are taken differently 
to clarify their peak structure in the same panel. 
See also \fig{ImReA-ph} obtained after a particle-hole transformation. 
}
\label{ImReA}
\end{figure}

In \fig{ImReA} we summarize ${\rm Im}\Sigma(\vk,\omega)$, ${\rm Re}\Sigma(\vk,\omega)$, 
and $A(\vk,\omega)$ as a function of $\omega$ for $\vk=(0, 0, \pi), (\pi,\pi, \pi)$, and 
$(\pi,0, \pi)$. Since $A(\vk,\omega)$ can form a peak when the following equation 
\be
\omega - \varepsilon_{\vk} - {\rm Re} \Sigma(\vk,\omega)=0
\label{QPcondition}
\ee
is fulfilled [see \eq{Akw}], we also add a line of $\omega - \varepsilon (\vk)$ in \fig{ImReA}.  
The line crosses the curve of ${\rm Re}\Sigma(\vk,\omega)$ typically
at three points for each $\vk$. 
One is very close to $\omega=0$ and yields the quasiparticle dispersion 
as already shown in \fig{QPdispersion}. 
The spectral function $A(\vk,\omega)$ exhibits a sharp peak there. 
The second one is close to the energy where ${\rm Im}\Sigma(\vk,\omega)$ has a peak, 
leading to a strong damping. As a result, $A(\vk,\omega)$ yields a very broad structure.  
The third one appears deeply below the Fermi energy, i.e., around $\omega/t \approx -4$, $-7$, and $-5$ 
in Figs.~\ref{ImReA}(a)-\ref{ImReA}(c), respectively, 
where ${\rm Im}\Sigma (\vk,\omega)$ almost 
vanishes and thus $A(\vk,\omega)$ forms a very sharp peak.

In addition, there are several sub-peak structures in $A(\vk,\omega)$.   
In $\omega>0$, ${\rm Im}\Sigma(\vk,\omega)$ can form a structure in 
$1 \lesssim \omega/t \lesssim 2.5$ as already explained in \fig{ImSig-asym}, which 
yields a small peak structure of ${\rm Re}\Sigma(\vk,\omega)$ there. 
The resulting value of $| \omega - \varepsilon (\vk) - {\rm Re}\Sigma(\vk,\omega) |$ 
[see \eq{QPcondition}] is suppressed, leading to the sub-peak structure of $A(\vk,\omega)$. 
In particular, the peak around $\omega/t \approx 1$ is pronounced in Figs.~\ref{ImReA}(b) and \ref{ImReA}(c).

\begin{figure}[ht]
\centering
\includegraphics[width=9cm]{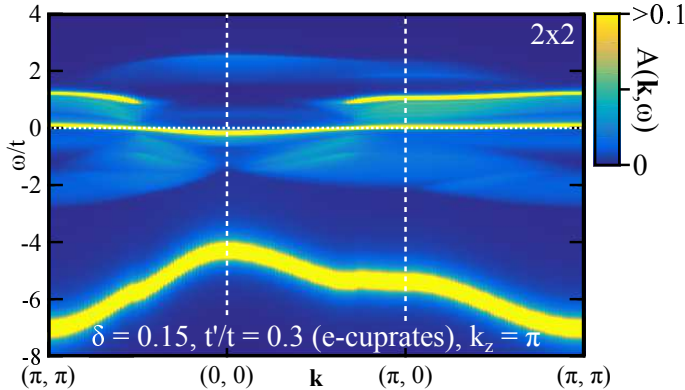}
\caption{(Color online) 
Intensity map of $A(\vk,\omega)$ along the direction $(\pi,\pi)$-$(0,0)$-$(\pi,0)$-$(\pi,\pi)$; 
$k_z$ dependence is weak and $k_z=\pi$ is taken as a representative value. 
See also \fig{Akw-ph}(b) obtained after a particle-hole transformation. 
}
\label{Akw-map}
\end{figure}

Figure~\ref{Akw-map} shows the intensity map of $A(\vk,\omega)$   
in the plane of $\vk$ and $\omega$. 
The quasiparticle band is realized around $\omega=0$. 
While it might look nondispersive in the energy scale in \fig{Akw-map}, 
it does disperse in a scale of $J$ $(=0.3t)$ as shown in \fig{QPdispersion}. 
A sharp dispersion in $-7 \lesssim \omega/t \lesssim -4$ is an emergent band arising from 
the coupling to the on-site charge fluctuations. 
In the positive energy region, one clear dispersion emerges 
around $\omega/t \approx 1$ especially along $(\pi,0)$-$(\pi,\pi)$-$(\pi/2,\pi/2)$ direction.   
The origin of this band lies in the acousticlike and optical plasmons around $\vq=(0,0,q_z)$ 
and is pronounced around $\vk=(\pi,0,k_z)$ and $(\pi,\pi,k_z)$, 
as we have explained in the context of Table~\ref{selection-rule}. 
The clear separation of this emergent band around $\omega/t \approx 1$ from the main quasiparticle band 
around $\omega \approx 0$ is due to the relatively large energy scale of plasmons around $\vq=(0,0,q_z)$. 
If we discard the plasmons by replacing the long-range Coulomb interaction 
with a short-range one, the charge excitations are characterized by a zero-sound mode 
around $\vq=(0,0,q_z)$ (see Ref.~\onlinecite{greco17}), which becomes 
gapless at $\vq=(0,0,0)$. In this case an emergent band corresponding to that around 
$\omega/t  \approx 1$ tends to merge into the main 
quasiparticle band (see \fig{Akw-map-SR}).

There are also weak and fine structures in \fig{Akw-map}. 
Two weak bandlike features emerge in $-1 \lesssim \omega/t \lesssim 0$ 
on both sides of  $\vk = (0,0,k_z)$. 
These weak structures are due to the peak structure of Im$\Sigma(\vk,\omega)$ 
as explained in \fig{ImReA}, yielding incoherent bands there. 
This incoherency comes from the coupling to the plasmons. 
Around $\omega/t \approx 2$ and $\vk \approx (0,0,k_z)$, we also have a weak structure, 
which is barely visible in \fig{Akw-map}. This incoherent band originates from the coupling 
to charge excitations around $\vq \sim (\pi,\pi, q_z)$.

\begin{figure}[ht]
\centering
\includegraphics[width=8.5cm]{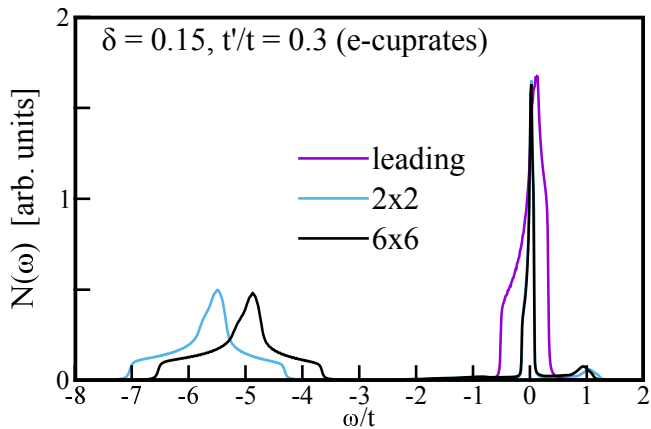}
\caption{(Color online) 
The density of states for three difference cases: 
one is at leading order in the large-$N$ theory [\eq{xik}]  
and the other two are $2\times 2$ and $6\times 6$, where the self-energy effect 
is taken into account from on-site charge fluctuations and both on-site charge and bond-charge 
fluctuations, respectively.  
The results for $2\times 2$ and $6\times 6$ almost overlap with each other 
around $\omega/t=0$ and $1$.  
See also \fig{Akw-ph}(d) obtained after a particle-hole transformation. 
}
\label{dos}
\end{figure}

In \fig{dos}, we show how the density of states is renormalized by electron correlation effects in 
the $t$-$J$ model. 
Our original band \eq{xik} is already renormalized at leading order 
as seen in \fig{QPdispersion}. 
Hence the density of states is limited to a narrow energy window in $-0.6 \lesssim \omega/t \lesssim 0.5$. 
The coupling to the on-site charge fluctuations ($2 \times 2$ in \fig{dos})  then splits 
the density of states into three. 
The resulting energy window of the density of states around the Fermi energy ($\omega=0$) shrinks 
substantially and two states are newly generated around $\omega/t \approx 1$ and $-5.5$.

\subsection{Effect of bond-charge fluctuations and doping dependence} 
\begin{figure}[ht]
\centering
\includegraphics[width=9cm]{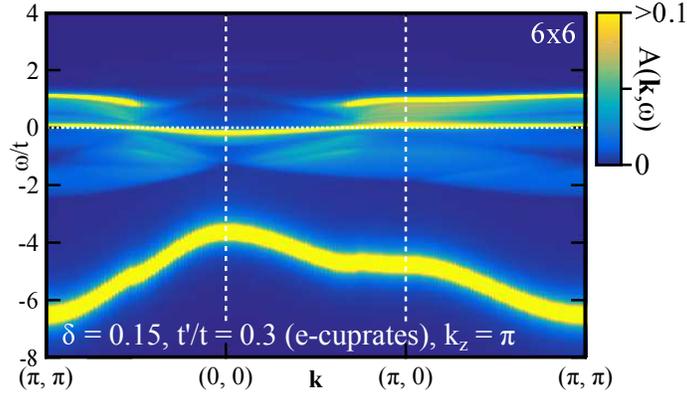}
\caption{(Color online) 
Intensity map of $A(\vk,\omega)$ along the direction $(\pi,\pi)$-$(0,0)$-$(\pi,0)$-$(\pi,\pi)$ 
in the presence of both on-site charge and bond-charge fluctuations. 
$k_z$ dependence is weak and $k_z=\pi$ is taken as a representative value. 
See also \fig{Akw-ph}(c) obtained after a particle-hole transformation. 
}
\label{Akw-all-map}
\end{figure}

So far we have restricted the summation of $a,b$ in \eq{ImSig} to $a,b=1,2$, that is, 
we have focused on the effect of the on-site charge fluctuations, which yield plasmons \cite{greco16,bejas17,greco19,greco20,nag20}. 
As emphasized in Ref.~\onlinecite{bejas17}, charge excitations in the $t$-$J$ model 
contain not only the on-site charge fluctuations but also bond-charge fluctuations. 
The bond-charge fluctuations are incorporated by allowing the indices $a$ and $b$ up to 6 in \eq{ImSig}. 
The obtained spectral function $A(\vk,\omega)$ is shown in \fig{Akw-all-map}. 
A comparison with \fig{Akw-map} 
demonstrates that essentially the same results are obtained even if the bond-charge fluctuations are 
taken into account. 
A close inspection reveals a suppression of an incoherent band around $\omega/t \approx 2$ 
and $\vk \approx (0, 0, k_z)$ and the resulting band becomes invisible in \fig{Akw-all-map}.   
It can be said that the one-particle excitation spectrum  $A(\vk,\omega)$ 
is mainly controlled by the on-site charge fluctuations and the effect of the bond-charge fluctuations 
is minor. 

This insight also applies to the density of states shown in \fig{dos}. It is interesting that 
the bond-charge fluctuations impact the density of states in a high-energy region 
around $\omega/t \approx -5$ more than that in $-0.5 \lesssim \omega /t  \lesssim 1$, 
although the typical energy scale of the bond-charge excitations is around $J$. 

\begin{figure}[ht]
\centering
\includegraphics[width=8cm]{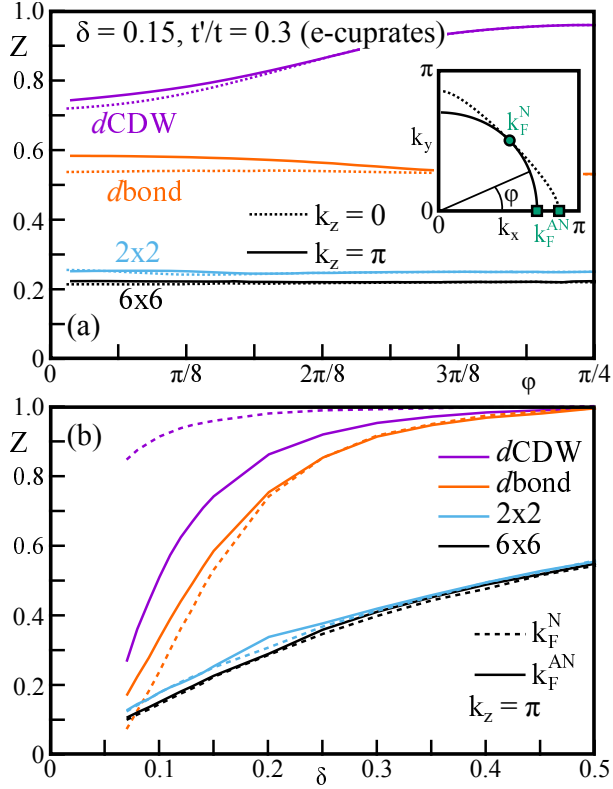}
\caption{(Color online) 
(a) Quasiparticle weight $Z(\varphi)$ along the Fermi surface 
for $k_z=0$ and $\pi$ at $\delta=0.15$. 
The momentum is measured by the angle $\varphi$ defined in the inset. 
Four different cases are plotted: 
i) $6 \times 6$, where full charge fluctuations are included, 
ii) $2 \times 2$, where only on-site charge fluctuations are considered, 
iii) $d$bond and iv) $d$CDW, where fluctuations associated with 
$d$bond and $d$CDW are considered, respectively. 
In the inset, Fermi surfaces shown by solid and dotted lines correspond to 
$k_z=\pi$ and $0$, respectively. 
(b) Doping dependence of $Z(\varphi)$ at 
$\varphi=0$ and $\pi/4$ 
for four different cases. The $d$CDW and $d$bond are stabilized below 
$\delta_c \approx 0.07$ and $0.06$, respectively. 
After the particle-hole transformation discussed in Sec.~III F, 
the results stay intact except that the momentum in the inset in (a) should be 
transformed as $(0,0) \rightarrow (\pi,\pi)$ and $(\pi,\pi) \rightarrow (0,0)$.
}
\label{Z-summary}
\end{figure}

Bond-charge fluctuations become stronger when the system is tuned closer to 
a bond-charge instability. However, their effect on the electron self-energy is still minor compared with 
the on-site charge fluctuations. 
To demonstrate this, we study how the quasiparticle weight is renormalized via coupling to 
charge fluctuations. 
In the present theory with $\Gamma_{\rm ch}/t=0.03$ [\eq{gamma-ch}], 
$d$-wave bond-charge ($d$bond) order \cite{yamase00a,yamase00b,metzner00,metzner03,yamase04b,metlitski10,holder12} 
and flux order \cite{affleck88a,morse91,cappelluti99},   
the so-called $d$-wave charge density wave ($d$CDW) \cite{chakravarty01},  
are stabilized below $\delta_c \approx 0.06$ and $0.07$, respectively, at $T=0$. 
The value of $\delta_c$ depends strongly on a choice of $\Gamma_{\rm ch}$, and Refs.~\onlinecite{bejas12} 
and \onlinecite{bejas14} studied charge instabilities for $\Gamma_{\rm ch}/t=0.01$. 
Since the quasiparticle weight can depend on the Fermi momentum, we first show its typical dependence 
[\fig{Z-summary}(a)] by setting $\delta=0.15$, which is away from the bond-charge instabilities, 
and then clarify its doping dependence afterward [\fig{Z-summary}(b)]. 

In \fig{Z-summary}(a), the momentum dependence of $Z(\varphi)$ is presented along the Fermi surface; 
$\varphi$ is defined in the inset. 
$Z(\varphi)$ associated with $d$CDW is close to unity at $\varphi=\pi/4$ and is reduced to around $0.7$ upon 
approaching $\varphi=0$, showing the anisotropy of $Z(\varphi)$ with a $d$-wave symmetry. 
$Z(\varphi)$ associated with $d$bond becomes smaller than the case of $d$CDW 
and is almost isotropic. 
This dependence might seem peculiar because it is not characterized by $d$-wave symmetry 
although $d$bond contains $d$-wave in its terminology. 
This puzzle is easily resolved by noting that $d$bond fluctuations 
have low-energy spectral weight around the in-plane momentum $(0.8\pi, 0.8\pi)$; 
see Fig.~4(a) in Ref.~\onlinecite{bejas17}. 
In this case, the $d$-wave form factor enhances the contribution from the nodal region 
around $\vk = (\pi/2, \pi/2, k_z)$ more than the antinodal region around $\vk=(\pi,0, k_z)$ and $(0,\pi, k_z)$. 
But it depends on the choice of parameters whether $Z$ at $\varphi=\pi/4$ 
eventually becomes larger than $Z$ at $\varphi=0$. 

On the other hand, $Z(\varphi)$ from the on-site charge fluctuations [$2 \times 2$ in \fig{Z-summary}(a)] 
is isotropic along the Fermi surface 
and becomes around $0.25$, much smaller than $Z$ from $d$CDW and $d$bond. 
In addition, $Z(\varphi)$ from $2 \times 2$ almost reproduces $Z(\varphi)$ from 
all charge fluctuations denoted by $6 \times 6$ in \fig{Z-summary}(a). 
This clearly indicates that the momentum dependence of $Z(\varphi)$ 
from $d$bond and $d$CDW is fully smeared out by the 
isotropic contribution from the on-site charge fluctuations.  

We also plot $Z(\varphi)$ for different values of $k_z$ in \fig{Z-summary}(a). 
$k_z$ dependence is visible for $d$CDW and $d$bond, but is rather weak. 
On the other hand, it is negligible for $2 \times 2$ and $6 \times 6$.    

The doping dependence of $Z(\varphi)$ is shown in \fig{Z-summary}(b) at two positions 
$\varphi=0$ and $\pi/4$ for different charge fluctuations. 
With decreasing doping rate, $Z(\varphi)$ at $\varphi=0$ for $d$CDW is substantially 
suppressed whereas $Z(\varphi)$ at $\varphi=\pi/4$ stays at a value near unity. 
The anisotropy of $Z(\varphi)$ is pronounced upon approaching the $d$CDW instability. 
The value of $Z(\varphi)$ for $d$bond decreases at both 
$\varphi=0$ and $\pi/4$ almost similarly with decreasing $\delta$ 
and their values become smaller than those for $d$CDW. 

For the purely on-site charge fluctuations denoted by $2 \times 2$ in \fig{Z-summary}(b), 
$Z(\varphi)$ is essentially isotropic in a whole doping region. 
Note that $Z(\varphi)$ decreases monotonically with decreasing $\delta$, 
although the number of carriers is decreased. 
The values of $Z(\varphi)$ for the full charge fluctuations denoted by $6 \times 6$  
are almost the same as those for on-site charge fluctuations. 
This clearly indicates that the effect of the on-site charge fluctuations 
on the electron self-energy is dominant over the bond-charge 
fluctuations such as $d$bond and $d$CDW in a whole doping region.

\subsection{No pseudogap feature} 
\begin{figure}[th]
\centering
\includegraphics[width=8cm]{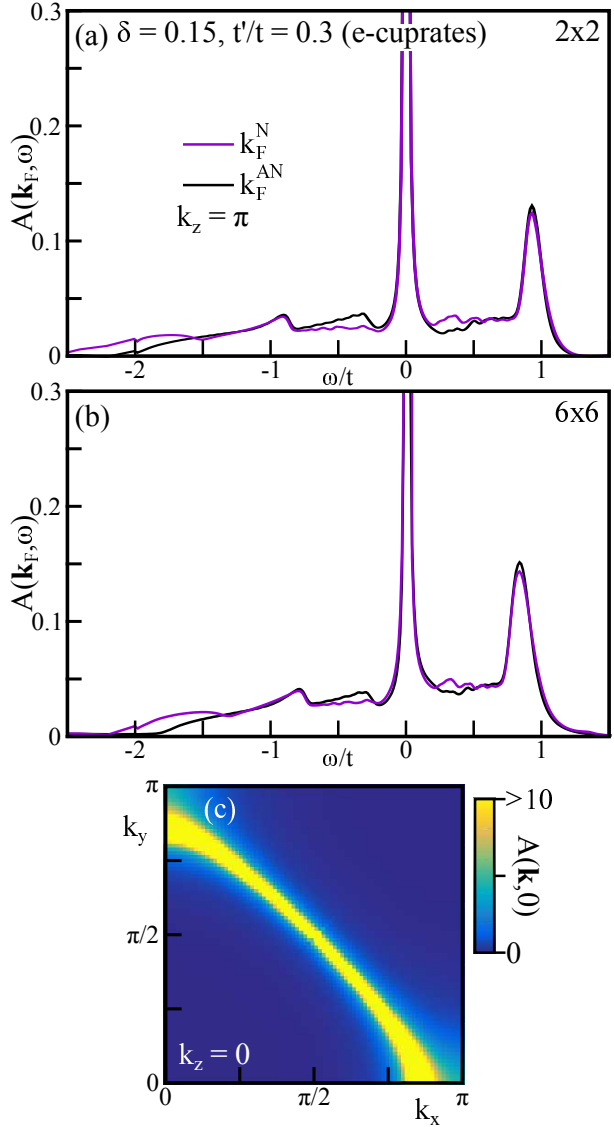}
\caption{(Color online) 
(a), (b) Spectral function $A(\vk_F,\omega)$ as a function of $\omega$ at the Fermi momentum; 
$k_z=\pi$ is taken as a representative one. 
$\vk_{F}^{\rm N}$ and $\vk_{F}^{\rm AN}$ are the Fermi momenta 
at the nodal $(\varphi=\pi/4)$ and antinodal $(\varphi=0)$ points, respectively. 
In (a) only on-site charge fluctuations (denoted by $2 \times 2$) are considered whereas 
the effect of both on-site charge and bond-charge fluctuations ($6 \times 6$) are taken 
into account in (b). 
(c) Intensity map of $A(\vk,0)$ at the Fermi energy in the first quadrant of the Brillouin zone 
for $k_z=0$ as a representative one. See also Figs.~\ref{Akw-ph}(e) and \ref{Akw-ph}(f) obtained 
after a particle-hole transformation.}
\label{pseudogap}
\end{figure}

Figures~\ref{QPdispersion} and \ref{dos}  imply that charge fluctuations do not lead to a pseudogap feature. 
To confirm this implication, we show in \fig{pseudogap}(a) the spectral function at the nodal and 
antinodal Fermi momenta by focusing on the on-site charge fluctuations. 
Two pronounced peaks are realized: one is the quasiparticle peak at $\omega=0$ and 
the other corresponds to the emergent incoherent band near $\omega \approx 1t$. 
This incoherent band is due to the coupling to the plasmons, as already explained in Sec. III B. 
There is a weak structure in $-1 \lesssim \omega/t \lesssim 0$, which is also recognized 
in the corresponding energy in \fig{Akw-map}. There is little difference 
between the nodal and antinodal Fermi momenta, 
demonstrating the isotropic renormalization effect from the on-site charge fluctuations. 

The spectral function shown in \fig{pseudogap}(b) is obtained by taking full charge fluctuations into account, 
namely both on-site charge and bond-charge fluctuations. 
A comparison with \fig{pseudogap}(a) shows that even though the typical energy scale of the 
bond-charge fluctuations is $J$, much lower than the on-site charge fluctuations, 
the effect of the bond-charge fluctuations are minor and no appreciable changes 
occur even around the Fermi energy. 

Figure~\ref{pseudogap}(c) is the intensity map of $A(\vk,0)$ at the Fermi energy 
in the first quadrant of the Brillouin zone. There is strong spectral weight 
entirely along the Fermi surface. 
A close look at \fig{pseudogap}(c) reveals that the spectrum becomes sharpest at the 
nodal direction and broadest at the antinodal region. This is due to the band structure effect, 
i.e., the proximity to the saddle points at $(\pi,0)$ and $(0,\pi)$ in the electron dispersion [\eq{Epara}]. 
Figures~\ref{QPdispersion} and \ref{dos} together with \fig{pseudogap} confirm 
no indication of a pseudogap feature in the presence of the quantum charge 
fluctuations in the $t$-$J$ model. 

We checked that the essential features in \fig{pseudogap} 
do not depend on doping (for $\delta > \delta_c$). 
Hence our Fermi surface 
is always large 
even though the system gets closer to a bond-charge instability at a lower doping 
and the spectral weight is suppressed substantially (\fig{Z-summary}). 

\subsection{Results for h-cuprates}
So far we have presented results which can be applicable to e-cuprates. 
For parameters appropriate for h-cuprates, we obtain very similar results. 
Nonetheless, we think it worthwhile presenting the corresponding results for h-cuprates and 
clarifying the meaning of {\it very similar results}, because 
h-cuprates are studied more than e-cuprates as a general trend. 
Following Ref.~\onlinecite{greco19}, we choose $t'/t=-0.2$ 
and the same values as those for e-cuprates 
for the rest of the parameters such as $t_z$, $V_c$, $\alpha$, $J$, and $\delta (=0.15)$.  

\begin{figure}[th]
\centering
\includegraphics[width=8cm]{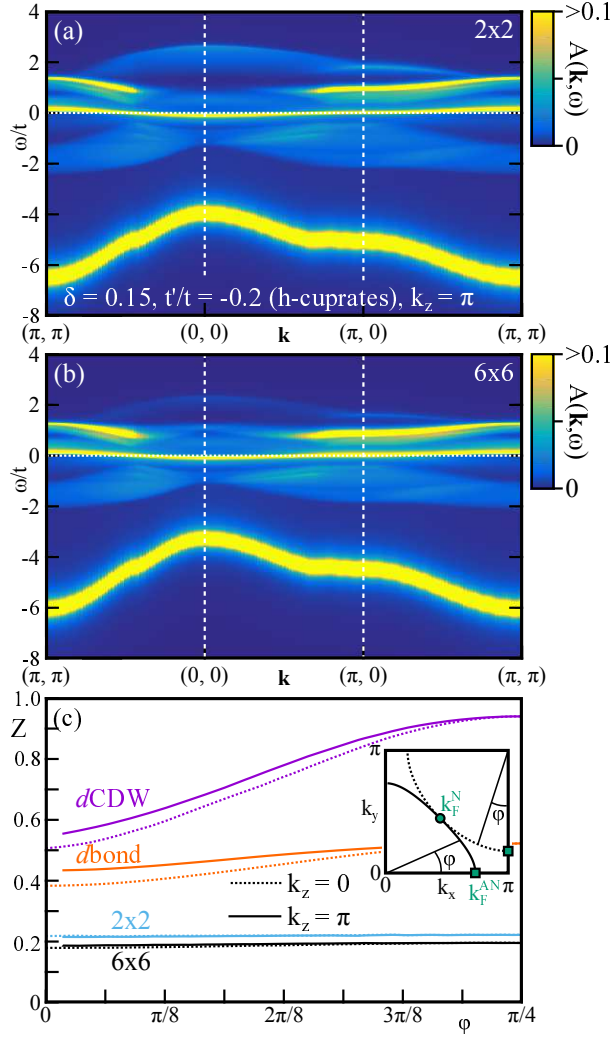}
\caption{(Color online) 
Results for parameters appropriate for h-cuprates at $\delta=0.15$. 
(a) Intensity map of $A(\vk,\omega)$ in the presence of the on-site charge fluctuations 
along the direction $(\pi,\pi)$-$(0,0)$-$(\pi,0)$-$(\pi,\pi)$; 
$k_z$ dependence is weak and $k_z=\pi$ is taken as a representative value. 
(b) Intensity map of $A(\vk,\omega)$ in the presence of both on-site charge 
and bond-charge fluctuations. 
(c) Quasiparticle weight $Z(\varphi)$ along the Fermi surface for four cases: 
$6 \times 6$ where full charge fluctuations are included, 
$2 \times 2$ where only on-site charge fluctuations are considered, and 
$d$bond and $d$CDW where fluctuations associated with 
$d$bond and $d$CDW are considered, respectively. 
The results for e-cuprates corresponding to (a)-(c) 
are shown in Figs.~\ref{Akw-map}, \ref{Akw-all-map}, and \ref{Z-summary}(a), respectively. 
}
\label{h-cuprate}
\end{figure}

Figure~\ref{h-cuprate}(a) is the intensity map of $A(\vk,\omega)$ due to the coupling to 
the on-site charge fluctuations and is practically the same as \fig{Akw-map}. 
This result does not change much even if we include the bond-charge fluctuations as shown in 
\fig{h-cuprate}(b) and is very similar to \fig{Akw-all-map}.  
The electron self-energy is dominated by the on-site charge fluctuations, the same conclusion 
as that for e-cuprates. 

Figure~\ref{h-cuprate}(c) shows the quasiparticle weight $Z(\varphi)$ along the Fermi surface.
This result is similar to the corresponding result for e-cuprates 
shown in \fig{Z-summary}(a). That is, the result for $2 \times 2$ is almost the same as 
that for $6 \times 6$, indicating that the effect of on-site charge fluctuations is dominant 
and $Z(\varphi)$ becomes almost isotropic along the Fermi surface. 
Since the present doping rate $\delta=0.15 $ is close to the critical doping rate 
of the $d$CDW instability ($\delta_c \approx 0.12$), the value of $Z(\varphi)$ is reduced 
more than that in \fig{Z-summary}(a) upon approaching $\varphi=0$. 
While the critical doping rate of the $d$bond instability is almost the same as that for e-cuprates, 
the value of $Z(\varphi)$ away from $\varphi = \pi/4$ and consequently 
its $\varphi$ dependence for $d$bond become somewhat different from that in \fig{Z-summary}(a). 
This reflects the difference of the band structure. 
In fact, the $\varphi$ dependence of $Z$ for $d$bond is rather sensitive to band parameters 
and a value of $Z$ at $\varphi=0$ can become larger or smaller than that at $\varphi=\pi/4$. 
For the present parameters for h-cuprates, the inner Fermi surface becomes electronlike for $k_z=\pi$; 
see the inset in \fig{h-cuprate}(c). 
It is remarkable that the difference of the Fermi surface topology between $k_z=0$ and $\pi$ 
does not produce appreciable changes of $Z(\varphi)$ not only for on-site charge fluctuations but also 
for bond-charge fluctuations in \fig{h-cuprate}(c). 

\begin{figure}[th]
\centering
\includegraphics[width=8cm]{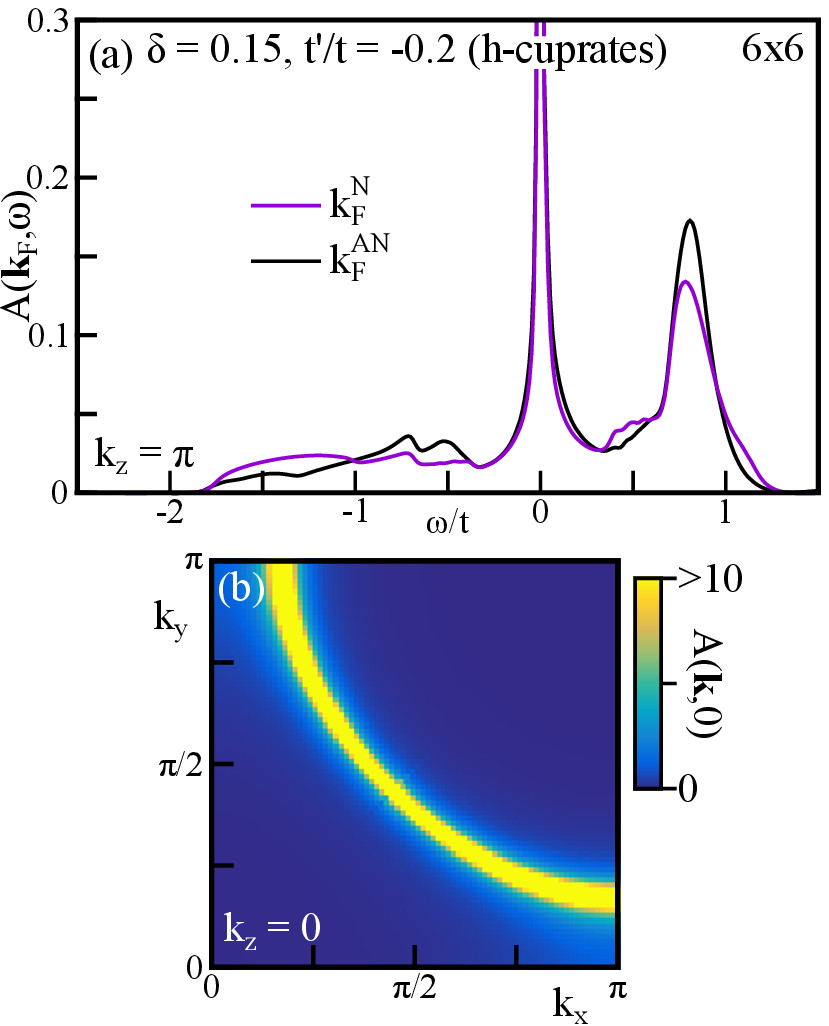}
\caption{(Color online) 
Results for h-cuprates at $\delta=0.15$. 
(a) Spectral function $A(\vk_{F},\omega)$ as a function of $\omega$ at the Fermi momentum. 
$\vk_{F}^{\rm N}$ and $\vk_{F}^{\rm AN}$ are the Fermi momenta 
at the nodal $(\varphi=\pi/4)$ and antinodal $(\varphi=0)$ points. 
(b) Intensity map of $A(\vk,0)$ at the Fermi energy in the first quadrant of the Brillouin zone. 
Corresponding results to e-cuprates are given in 
Figs.~\ref{pseudogap}(b) and \ref{pseudogap}(c). 
}
\label{h-cuprate-2}
\end{figure}

Figure~\ref{h-cuprate-2}(a) is the spectral function $A(\vk_{F},\omega)$  as a function of $\omega$ 
at the nodal ($\varphi=\pi/4$) and the antinodal ($\varphi=0$) Fermi momenta, 
and \fig{h-cuprate-2}(b) is the intensity map of $A(\vk,0)$ at the Fermi energy. 
These results should be compared with Figs.~\ref{pseudogap}(b) and \ref{pseudogap}(c). 
No appreciable differences are recognized. We can conclude that the quantum charge fluctuations 
do not lead to a pseudogap feature even for h-cuprates.

\subsection{Particle-hole transformation of results for e-cuprates}
There is a special aspect of the $t$-$J$ model in a study of the one-particle properties for e-cuprates. 
That is, the e-cuprates are analyzed in terms of the {\it hole} picture  in the $t$-$J$ model. 
This is because  the $t$-$J$ model is defined in the restricted Hilbert space 
where the double occupancy of electrons is forbidden. 
Given that experimental data in e-cuprates are analyzed in the {\it particle} picture, 
it should be more transparent to present our results in the same picture.   

We perform a particle-hole transformation in momentum space: 
$\tilde{c}_{\vk \sigma} \rightarrow \tilde{c}_{\vk+\vQ \sigma}^{\dagger}$ and 
$\tilde{c}_{\vk \sigma}^{\dagger} \rightarrow \tilde{c}_{\vk+\vQ \sigma}$ with $\vQ=(\pi,\pi,0)$ (Ref.~\onlinecite{misc-ph}). 
This yields the following changes: 
\bea
&& \epsilon_{\vk} \rightarrow - \epsilon_{\vk+\vQ} \,, \\
&&{\rm Re} \Sigma (\vk, \omega) \rightarrow -{\rm Re} \Sigma (\vk+\vQ, -\omega) \,, \\
&&{\rm Im} \Sigma (\vk, \omega) \rightarrow {\rm Im} \Sigma (\vk+\vQ, -\omega) \,, \\
&&A(\vk, \omega)  \rightarrow A(\vk+\vQ, -\omega) \,.
\eea
The charge correlation function does not change. 
Figures~\ref{QPdispersion}, \ref{Akw-map}, \ref{dos}, \ref{Akw-all-map}, and \ref{pseudogap} are transformed to 
those summarized in \fig{Akw-ph}.  
Results corresponding to 
Figures~\ref{ImSig-asym} and \ref{ImReA} are presented in Appendix~B. 
Note that this kind of transformation is not necessary for h-cuprates.

\begin{figure}[ht]
\centering
\includegraphics[width=16cm]{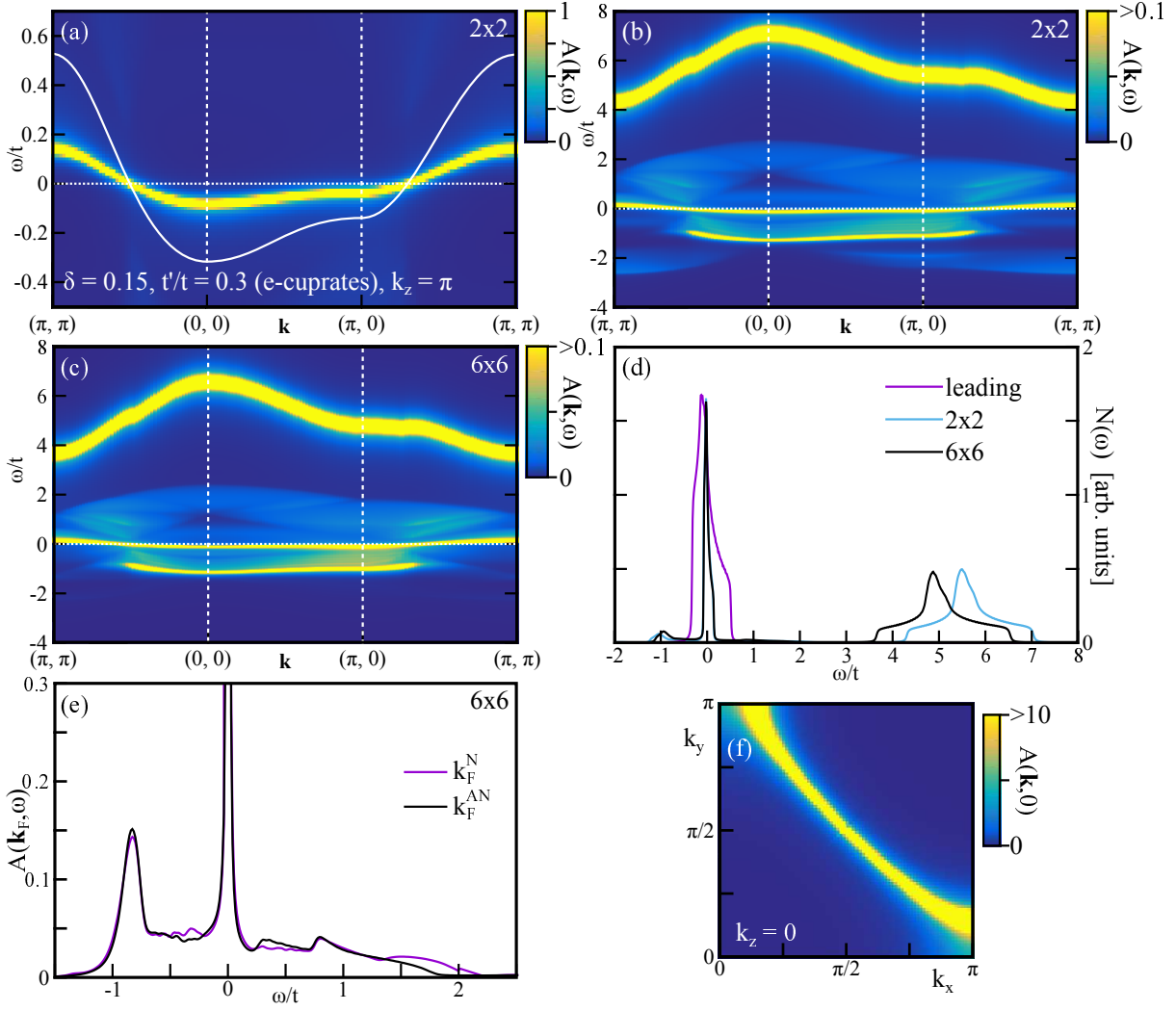}
\caption{(Color online) 
Intensity map of $A(\vk,\omega)$ along the direction $(\pi,\pi)$-$(0,0)$-$(\pi,0)$-$(\pi,\pi)$ 
after the particle-hole transformation 
(a) around the Fermi energy corresponding to \fig{QPdispersion}, and 
(b) and (c) in a larger energy region corresponding to \fig{Akw-map} and \ref{Akw-all-map}, respectively.  
(d) The density of states after the particle-hole transformation of \fig{dos}. 
(e) The spectral function $A(\vk_F, \omega)$ at the nodal and antinodal Fermi momenta.  
(f) Intensity map of $A(\vk, 0)$ at the Fermi energy in the first quadrant of the 
Brillouin zone. (e) and (f) are obtained after the particle-hole transformation of Figs.~\ref{pseudogap}(b)
and \ref{pseudogap}(c), respectively. 
}
\label{Akw-ph}
\end{figure}

\subsection{Absence of long-range Coulomb interaction} 
\begin{figure}[ht]
\centering
\includegraphics[width=9cm]{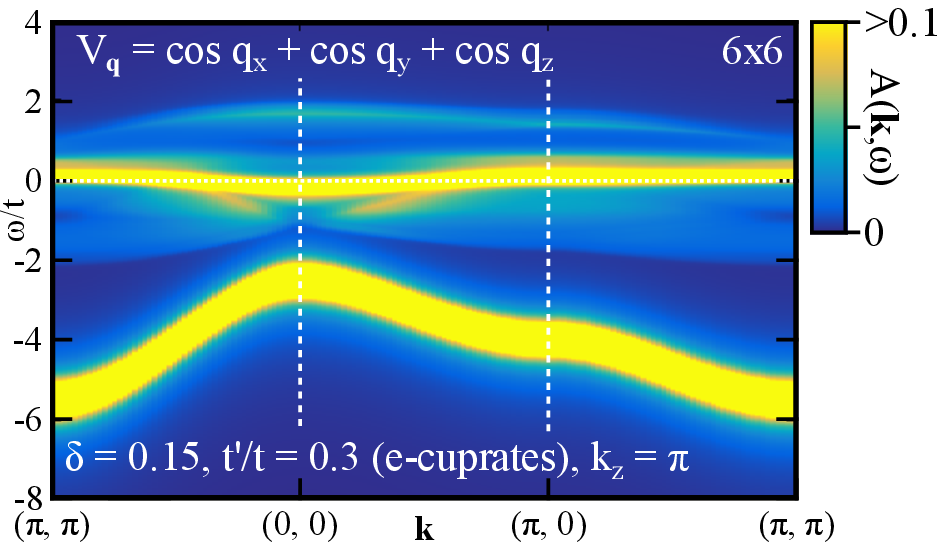}
\caption{(Color online) 
Intensity map of $A(\vk,\omega)$ in the absence of the long-range Coulomb interaction 
along the direction $(\pi,\pi)$-$(0,0)$-$(\pi,0)$-$(\pi,\pi)$ 
in the {\it hole} picture; 
both on-site charge and bond-charge fluctuations are taken into account. 
$k_z$ dependence is weak and $k_z=\pi$ is taken as a representative value. 
The corresponding result in the {\it particle} picture is presented in Appendix~B. 
}
\label{Akw-map-SR}
\end{figure}

We have considered the long-range Coulomb interaction [\eq{LRC}].  
What happens if we replace it with a short-range Coulomb interaction? 
As a typical short-range Coulomb interaction, we may consider 
\be
V(\vq)= V_1 (\cos q_x + \cos q_y) + V_2 \cos q_z \,.
\label{SRC}
\ee
The charge excitation spectrum for $V_2=0$ and $t_z=0$ was already shown in 
Ref.~\onlinecite{greco17}, where the zero-sound mode is realized as collective excitations, 
which are gapless at $\vq=(0,0,q_z)$ independent of $q_z$. 
In the present layered model with a finite $t_z$, the zero-sound mode acquires  a gap 
at $\vq=(0,0,q_z)$ for $q_z \ne 0$, but remains gapless for $q_z=0$; 
see Ref.~\onlinecite{greco19} for further details. 
We have chosen $V_1=V_2= 1$ and computed the intensity map of the one-particle spectral 
function. 

The obtained result is shown in \fig{Akw-map-SR}, which should be compared with 
\fig{Akw-all-map}, where the long-range Coulomb interaction is considered. 
The crucial differences appear in the two incoherent bands above the Fermi energy. 
The band around $\omega/t \approx 1$ seen in \fig{Akw-all-map} is now pushed down toward 
the main quasiparticle band 
and is hardly recognized in \fig{Akw-map-SR}. 
The other incoherent band around $\omega/t \approx 2$ 
becomes clearer than Fig.~\ref{Akw-all-map} (see also \fig{Akw-map}), 
where  the corresponding band does form around 
$\omega/t  \approx 2$, but is invisible in that intensity scale. 
Minor differences are also recognized. 
We now have a clear {\it mustache} in \fig{Akw-map-SR} just below the quasiparticle 
dispersion on both sides of $\vk=(0, 0, k_z)$, which comes from coupling to the zero-sound mode. 
The sharp side band deeply below the Fermi energy is robust. 
Quantitatively it is pushed up and becomes closer to the main quasiparticle band.

\subsection{$\boldsymbol{t'}$ dependence}
The band parameter $t'$ is often utilized to describe a material dependence in the cuprate family. 
In fact,  the tendency to various bond-charge-order instabilities depends on the value of $t'$ 
and in particular the sign change of $t'$ leads to a big difference \cite{bejas12, bejas14}. 
However, the effect of bond-charge fluctuations turns out to be very weak on the electron self-energy. 
Moreover, essential features of on-site charge fluctuations including plasmons are not altered 
by a different choice of $t'$ unless one is interested in fine fitting to experimental data \cite{nag20}. 
Hence the results obtained in the present paper do not depend much on a choice of $t'$. 
This is also recognized by noting similar results for both e- and h-cuprates; 
see Figs.~\ref{Akw-map}, \ref{Akw-all-map}, \ref{h-cuprate}(a) and \ref{h-cuprate}(b). 
We therefore expect that the present results manifest a generic feature of the spectral function 
when considering the electron self-energy from quantum charge fluctuations in the $t$-$J$ model.

\section{Discussions}
We have found that 
the electron self-energy is controlled mainly by the on-site charge fluctuations describing plasmons. 
This conclusion might seem counterintuitive because there are also bond-charge fluctuations, 
which are present in the low-energy region with the scale of $J$ and thus seem to 
contribute much to the self-energy especially near the Fermi energy. 
The point lies in the strength of the coupling between bond-charge fluctuations and electrons. 
The on-site charge fluctuations are described by the components 
$a,b=1,2$ and the bond-charge ones by $a,b=3$-$6$ in \eq{ImSig}. 
The vertex $h_a(\vk,\vq,\nu)$ with 
$a=3$-$6$ is proportional to $\Delta$ [see \eq{vertex-h}], whose typical value is 
around $0.025$-$0.03t$. 
Since the contribution from the bond-charge fluctuations is 
proportional to $\Delta^2$, 
its effect becomes much weaker than that from the on-site charge fluctuations. 

The factor of $\Delta^2$ can be generic to strongly correlated electron systems. 
In fact, similar features are recognized in 
slave-boson calculations \cite{wang91}, 
Kadanoff-Baym method for Hubbard operators \cite{greco01}, 
the equation of motion method for the Green's functions in terms of the Hubbard operators \cite{plakida14}, 
and the equations of motion for projected operators on the $t$-$J$ model \cite{prelovsek01}. 

The results for e-cuprates [Figs.~\ref{Akw-ph}(b) and \ref{Akw-ph}(c)] look 
different from those for h-cuprates [Figs.~\ref{h-cuprate}(a) and \ref{h-cuprate}(b)] in a usual {\it particle} picture. 
However, when we take a {\it hole} picture for e-cuprates (Figs.~\ref{Akw-map} and \ref{Akw-all-map}), 
those become essentially the same. 
This is because the obtained self-energy and the resulting one-particle spectral function 
are determined mainly by the coupling to the on-site charge fluctuations, which are almost the 
same for both e- and h-cuprates in the present theory \cite{greco19}. 
The major difference of charge excitations between e- and h-cuprates appears 
in the bond-charge fluctuations \cite{bejas14}, whose contribution to the self-energy is, however, 
minor even close to bond-charge instabilities at $T=0$. 
Given that the on-site charge excitations obtained in the present theory capture 
the experimental data in both e- and h-cuprates \cite{greco16,greco19,greco20,nag20}, 
the present paper implies that the obtained one-particle excitation spectra are universal in cuprates. 

We have shown that the coupling to plasmon excitations generates  
an incoherent  band around $1t$ in Figs.~\ref{h-cuprate}(a) and \ref{h-cuprate}(b). 
This emergent band can also be seen as 
the density of states generated in the corresponding energy region in \fig{dos}. 
In fact, early studies reported the peak around 1 eV above the Fermi energy 
by the inverse photoemission spectroscopy for h-cuprates \cite{drube89,wagener89}. 
The reported state can be our incoherent band generated by the coupling to plasmons 
since we can reasonably assume $t/2 \approx 500$ meV for cuprates 
(Refs.~\onlinecite{misc-factor2} and \onlinecite{hybertsen90}). 
The spectral weight of our incoherent band is enhanced along the direction $(\pi,0)$-$(\pi,\pi)$-$(\pi/2, \pi/2)$, 
which can be tested by the inverse ARPES. 
On the other hand, for e-cuprates, this incoherent band emerges below 
the Fermi energy along the direction $(\pi/2,\pi/2)$-$(0,0)$-$(\pi, 0)$ 
as shown in Figs.~\ref{Akw-ph}(b) and \ref{Akw-ph}(c) 
(in a usual {\it particle} picture). Given that plasmon excitations are now observed 
in experiments \cite{hepting18,jlin20,singh20,nag20}, 
it is interesting to test the predicted band by performing ARPES for e-cuprates.  

This test is also important to highlight the role of 
the long-range Coulomb interaction in cuprates. 
Traditionally, short-range electron-electron interactions are presumed in most of studies on cuprates. 
In fact, there have been experimental reports recently 
which are in disfavor with plasmon excitations \cite{mitrano19,husain19}. 
In this context, our results without the long-range Coulomb interaction 
(Figs.~\ref{Akw-map-SR} and \ref{Akw-map-SR-ph}) are useful. 
They confirm that the emergent incoherent band around $|\omega/t| \approx 1$ 
[Figs.~\ref{h-cuprate}(a), \ref{h-cuprate}(b), Figs.~\ref{Akw-ph}(b), and \ref{Akw-ph}(c)] indeed comes from the coupling to plasmons, 
namely, through the long-range Coulomb interaction.

The on-site charge fluctuations also generate a band around $5t$ for e-cuprates 
[Figs.~\ref{Akw-ph}(b) and \ref{Akw-ph}(c)] and around $-5t$ h-cuprates [Figs.~\ref{h-cuprate}(a) and \ref{h-cuprate}(b)]. 
A very similar feature was also obtained in the two-mode variational Monte Carlo study of the $t$-$J$ model \cite{tan08}. 
It may not be easy to test this emergent band experimentally since the distinction from additional bands, 
not included in the present one-band $t$-$J$ model, is not straightforward in real materials.

As shown in \fig{Z-summary}, the quasi-particle weight $Z$ is suppressed dramatically by on-site 
charge fluctuations and becomes smaller with decreasing carrier density, 
although charge fluctuations themselves are driven by the doped carrier.  
The similar doping dependence of $Z$ was also obtained in the multiband Hubbard model 
in the local density-approximation in combination with the dynamical mean-field theory \cite{weber10}. 
The present paper suggests that charge fluctuations are the major contribution to yield the suppression 
of the quasiparticle weight 
in the doped Mott insulator even though magnetic fluctuations are expected strong near half-filling.  

It is well known that one of the most important issues in cuprates is the pseudogap phase \cite{keimer15}, which is clearly 
resolved in h-cuprates, while its existence is controversial in e-cuprates \cite{armitage10}.
As seen in Figs.~\ref{dos}, \ref{pseudogap}, and \ref{h-cuprate-2}, 
the quantum charge fluctuations do not produce a pseudogap feature, 
although the spectral width in Figs.~\ref{pseudogap}(c) and \ref{h-cuprate-2}(b) 
becomes sharpest at the nodal region and broadest at the antinodal region, which 
somewhat shares a $d$-wave feature, the same symmetry of the pseudogap. 
It seems too hasty to infer that the pseudogap may originate from some other degrees of freedom 
such as spin fluctuations. 
In fact, our calculations have been performed at $T=0$ and thus 
the charge fluctuations we have studied are purely quantum. 
At finite temperatures, classical fluctuations also contribute 
to the self-energy \cite{greco09,yamase12}. We leave this possibility to a future issue. 

A renormalization of the electron dispersion is related directly to the so-called kinks 
of the dispersion observed by ARPES. Mainly two different kinks are reported: 
low-energy (40-70 meV)  (Refs.~\onlinecite{kaminski01,lanzara01,johnson01,gromko03,zhou05,chang07,dmou17,valla20}) 
and high-energy ($\sim$ 0.3 eV) (Refs.~\onlinecite{graf07,xie07,valla07,meevasana07,chang07,zhang08}) kinks. 
The low-energy kink was discussed in terms of a coupling to magnetic resonance mode  \cite{kaminski01,johnson01,gromko03,eschrig00,carbotte11,dmou17}, spin fluctuations \cite{markiewicz07,valla20}, and phonons  \cite{lanzara01,zeyher01,zhou05,carbotte11}, and also even without invoking 
such a coupling to bosons \cite{byczuk07}, whereas the high-energy kink may come from 
spin fluctuations \cite{macridin07,kar11},  incoherent hole motion \cite{zemljic08}, 
string excitations \cite{manousakis07}, and some extrinsic effects \cite{inosov07}. 
Although we have performed a comprehensive analysis of the electron self-energy from 
all possible charge fluctuations included in the $t$-$J$ model at order of $1/N$, 
we did not obtain a clear kink structure associated with experimental data 
if we assume $t/2 \approx 500$ meV. 

The fact that we did not obtain a clear kink might seem peculiar because 
it is well recognized that the electron dispersion may exhibit a kink near a typical energy scale of 
bosonic fluctuations via their coupling to electrons. The reasons are twofold: energy scale and coupling strength. 
As shown in \fig{QPdispersion}, the quasiparticle dispersion is strongly renormalized by the coupling 
to charge fluctuations and is realized in $-0.2 \lesssim \omega/t \lesssim 0.1$. 
In such a low energy scale, it is only bond-charge fluctuations which could play a role 
because on-site charge fluctuations have energy much larger than the renormalized dispersion. 
However, as we have shown and discussed, the effect of bond-charge fluctuations is substantially weakened 
by the factor of $\Delta^2$ coming from the vertex $h_a(\vk, \vq, \nu)$ [see \eq{vertex-h}] and 
cannot generate a clear kink in the electron dispersion.

Our results shown in Figs.~\ref{Akw-map}, \ref{Akw-all-map}, and \ref{h-cuprate}(a) and \ref{h-cuprate}(b) 
indicate a large spectral weight around $\omega \sim -4t$ near $\vk=(0, 0, k_z)$, which was also  
obtained in finite temperature diagonalization of small clusters \cite{zemljic08}. 
Although Ref.~\onlinecite{zemljic08} discussed the high-energy kink \cite{graf07,xie07,valla07,meevasana07,chang07,zhang08} on the basis of such results, 
we do not because the absolute energy becomes different more than a factor of 4 
if we assume $t/2 \approx 500$ meV. 
Considering that our charge excitation spectra capture the experimental data, including 
quantitative aspects in many cases \cite{yamase15b,yamase19,greco19,greco20,nag20}, 
we think that a factor of 4 is too large as the energy difference. 
We hope that this different view from Ref.~\onlinecite{zemljic08} stimulates a further study of the high-energy kink, 
whose origin remains controversial and will lead to a better understanding of the nature of the self-energy in cuprates. 

Instead of a kink in the electron dispersion, we predict a {\it cascadelike} feature 
in the sense that the spectral weight of the quasiparticle dispersion {\it flows} to 
the incoherent band around $\omega/t \sim -1$ especially in {\it wide regions} around 
$\vk \approx  (\pi, 0, k_z)$ and $(\pi/2, \pi/2, k_z)$ for e-cuprates in Figs.~\ref{Akw-ph}(b) and \ref{Akw-ph}(c). 
This feature should not be confused with the ``waterfall" reported around $-(0.3 \sim 1)$ eV 
and $\vk \approx (0, 0, k_z)$ for h-cuprates \cite{graf07,xie07,valla07,meevasana07,chang07,zhang08}. 
We did not obtain such a waterfall clearly for h-cuprates; see Figs.~\ref{h-cuprate}(a) and \ref{h-cuprate}(b).

The effect of plasmons on the electron dispersion in h-cuprates was studied 
in Ref.~\onlinecite{markiewicz07a} in a phenomenological scheme. 
They reported that a coupling to plasmons generates 
shadow bands around 1.5 eV below and above the LDA band. 
This qualitative feature is shared with our results in that we have also obtained
bands generated around the energy $- 5t$ and $1t$ in  Figs.~\ref{h-cuprate}(a) and \ref{h-cuprate}(b). 
Furthermore, Ref.~\onlinecite{markiewicz07a} concluded that plasmons cannot explain 
the high-energy kink of the electron 
dispersion around $-(0.3 \sim 1)$ eV (Refs.~\onlinecite{graf07,xie07,valla07,meevasana07,chang07,zhang08}), 
but the plasmons strongly renormalize the electron dispersion. 
These insights are also shared with ours obtained in the present microscopic analysis  
of the self-energy. 

Although we do not share a view of the high-energy kink with Ref.~\onlinecite{zemljic08}, 
the present work is complementary to Ref.~\onlinecite{zemljic08}. 
Reference~\onlinecite{zemljic08} performed numerical analysis by using small clusters of 
the $t$-$J$ model, but did not consider the effect of plasmons. 
On the other hand, we focus on the charge sector and analyze all fluctuations including plasmons 
in the thermodynamic limit. 
Since both magnetic and charge fluctuations were involved in Ref.~\onlinecite{zemljic08}, 
it remained unclear what kind of fluctuations were actually relevant to the spectral function. 
The present paper then clarifies that strong electron correlations due to the on-site charge fluctuations 
including plasmons, not bond-charge nor magnetic ones, play a dominant role to form the electron spectral function. 
Reference~\onlinecite{zemljic08} obtained the imaginary part of the electron self-energy 
with strong particle-hole asymmetry. 
This asymmetry was understood as a consequence from the strong correlation effects due to the local constraint in 
the $t$-$J$ model in Ref.~\onlinecite{zemljic08}. 
This insight fully agrees with the present paper as we explain in the context of \fig{ImSig-asym}. 
While Ref.~\onlinecite{zemljic08} did not consider the long-range Coulomb interaction, 
we find that it leads to a side band around $| \omega/t |\approx 1$. 
In addition, we elucidate the quasi-particle weight, 
its doping dependence, and the interplay of on-site charge and bond-charge fluctuations.

Reference~\onlinecite{fleck01} studied the spectral function in the charge-stripe phase. 
While no  tendency toward charge stripes was obtained in the present theory \cite{bejas12,bejas14}, 
a certain bond-charge order has the same symmetry as stripes \cite{bejas12}. 
In this sense, we might mimic the effect of stripes on the spectral function and 
make an interesting comparison with Ref.~\onlinecite{fleck01}. 
However, as we have showed in the present paper, the effect of bond-charge fluctuations is 
minor on the electron self-energy.

We have employed a layered model to perform a realistic analysis of the self-energy 
from the charge excitations and considered the long-range Coulomb interaction 
in three-dimensional space [see \eq{LRC}].  This three-dimensional Coulomb interaction 
is important to take a plasmon contribution into account appropriately. 
Concerning the electron band structure, however, we find that essentially the same results 
are obtained for the electron self-energy even if we set $t_z=0$. 
In this sense, the detail of three dimensionality of the band is not crucial to the electron self-energy 
from charge fluctuations, at least at zero temperature. 

The oxygen degrees of freedom in cuprates are included effectively  
in the one-band $t$-$J$ model as the Zhang-Rice singlet \cite{fczhang88}  
and charge fluctuations at the oxygen sites are described as bond-charge fluctuations in the present theory. 
However, this does not mean that the effect of oxygens is included perfectly in the $t$-$J$ model. 
For example, if we wish to address the issue why Cu L-edge RIXS detects the plasmons in e-cuprates 
efficiently \cite{ishii05,wslee14,ishii14,dellea17,hepting18,jlin20} 
while it does not in h-cuprates \cite{ishii17, nag20}, 
we feel that the $t$-$J$ model is not enough and 
the oxygen degrees of freedom should be considered on an equal footing.

Finally, it is natural to ask whether our obtained results could stay if one studies 
the electron self-energy beyond the present approximation at the order of $1/N$. 
While this possibility is not easily addressed, we believe that the essential features may not be modified much 
on the basis of a comparison with numerical analysis of the $t$-$J$ (Refs.~\onlinecite{zemljic08} and \onlinecite{tan08}) 
and Hubbard \cite{weber10}  models as we have discussed above. 

\section{Concluding remarks}
The electron self-energy in cuprates was mainly studied by considering a coupling 
to spin fluctuations \cite{kaminski01,eschrig00,johnson01,gromko03,markiewicz07,macridin07,carbotte11,kar11,dmou17,valla20}
and phonons \cite{lanzara01,zhou05,carbotte11,zeyher01}.   
Compared with those studies, the effect of charge fluctuations on the electron self-energy 
has been studied much less \cite{markiewicz07a,zemljic08}.  
Given that high-$T_c$ superconductivity occurs upon doping of charged carriers \cite{keimer15} 
and the charge dynamics was revealed in $(\vq,\omega)$ space recently \cite{ishii05,ghiringhelli12,chang12,achkar12,wslee14,ishii14,da-silva-neto15,da-silva-neto16,dellea17, ishii17,hepting18,jlin20,singh20,nag20}, 
it is challenging to clarify how the actual charge dynamics impacts the one-particle properties, 
which will then deepen our understanding of the cuprate phenomenology. 
In this paper, we have performed a comprehensive analysis of the electron self-energy from 
quantum charge fluctuations in a realistic layered $t$-$J$ model, 
which can capture the charge excitation spectra observed in cuprates.
We have found that on-site charge fluctuations have a great 
impact on the electron self-energy. 
The quasiparticle weight is reduced significantly, 
which occurs almost isotropically along the Fermi surface. 
In addition, mainly two additional 
bands emerge. In particular, the emergent band 
around $\omega/t  \approx 1$ with strong intensity along $(\pi,0)$-$(\pi,\pi)$-$(\pi/2,\pi/2)$  
in Figs.~\ref{Akw-map}, \ref{Akw-all-map}, and \ref{h-cuprate} 
[see also Figs.~\ref{Akw-ph}(b) and \ref{Akw-ph}(c) where it occurs around $\omega/t \approx -1$ 
along  $(\pi/2,\pi/2)$-$(0,0)$-$(\pi,0)$]  
comes from the coupling to the plasmons \cite{greco16,greco19,greco20} 
recently identified by x-ray measurements \cite{hepting18,jlin20,singh20,nag20}. 
Surprisingly, the effect of low-energy bond-charge quantum fluctuations is minor even 
close to bond-charge instabilities and the electron self-energy is mainly controlled by 
the on-site charge fluctuations describing plasmons. 
Furthermore, the quantum charge dynamics does not produce a clear kink nor a pseudogap feature 
in the electron dispersion.

\acknowledgments
The authors thank M. Hepting, P. Horsch, A. M. Ole\'s, and R. Zeyher for fruitful comments. 
A.G. thanks the Max-Planck-Institute for Solid State Research in Stuttgart for hospitality and 
financial support.  H.Y. was supported by JSPS KAKENHI Grants No.~JP18K18744 
and No.~JP20H01856.



\appendix
\newpage
\section{Complete formalism} 
We start with the layered $t$-$J$ model with the long-range Coulomb interaction: 
\begin{equation}
H = -\sum_{i, j,\sigma} t_{i j}\tilde{c}^\dag_{i\sigma}\tilde{c}_{j\sigma} + 
\sum_{\langle i,j \rangle} J_{ij} \left( \vec{S}_i \cdot \vec{S}_j - \frac{1}{4} n_i n_j \right)
+ \frac{1}{2} \sum_{i,j } V_{ij} n_i n_j \, .
\label{tJV-A}  
\end{equation}
This model is defined in the restricted Hilbert space, where the double occupancy of electrons  
is prohibited at any lattice site. 
We can write the $t$-$J$ model in terms of nine Hubbard $\hat{X}$ operators \cite{hubbard63}: 
$\tilde{c}^\dag_{i \sigma}=\hat{X}_i^{\sigma 0}$, 
$\tilde{c}_{i \sigma}=\hat{X}_i^{0 \sigma}$, $S_i^+=\hat{X}_i^{\uparrow \downarrow}$, 
$S_i^-=\hat{X}_i^{\downarrow \uparrow}$, $n_i=\hat{X}_i^{\uparrow \uparrow}+\hat{X}_i^{\downarrow \downarrow}$, 
and $\hat{X}_i^{0 0}$ describing the number of doped holes; the $z$ component of the spin operator 
is described by $S_{i}^{z} = \frac{1}{2}(\hat{X}_{i}^{\uparrow \uparrow} -  \hat{X}_{i}^{\downarrow \downarrow})$.  
The operators $\hat{X}_i^{\sigma 0}$ and $\hat{X}_i^{0 \sigma}$ are called fermionlike, 
whereas the operators $\hat{X}_i^{\sigma \sigma'}$ and 
$\hat{X}_i^{00}$ are bosonlike. 
After introducing the chemical potential $\mu$, the resulting Hamiltonian becomes 
\begin{eqnarray}
&&H(X) = -\sum_{i, j, \sigma}\; t_{i j} \hat{X}_{i}^{\sigma 0}\hat{X}_{j}^{0\sigma} 
+ \frac{J}{2} \sum_{\bra i,j\ket, \sigma \sigma'} (\hat{X}_{i}^{\sigma \sigma'}\hat{X}_{j}^{\sigma' \sigma} - 
\hat{X}_{i}^{\sigma \sigma} \hat{X}_{j}^{\sigma' \sigma'}) 
 \nonumber\\
&&\hspace{15mm} + \frac{1}{2} \sum_{i,j, \sigma \sigma'} V_{ij} \hat{X}_{i}^{\sigma \sigma} \hat{X}_{j}^{\sigma' \sigma'}
-\mu\sum_{i,\sigma}\;\hat{X}_{i}^{\sigma \sigma} \,.
\label{eq:H} 
\end{eqnarray}

There are two major difficulties in the Hamiltonian (\ref{eq:H}): the complicated
commutation rules of the Hubbard operators \cite{hubbard63} and 
the absence of a small parameter. A popular method to 
handle the former difficulty is to use slave particles. 
For instance, in the slave-boson method \cite{kotliar88} 
the original fermionlike $\hat{X}^{0 \sigma}$ operator is written as 
$\hat{X}^{0\sigma} = \hat{b}^{\dagger} \hat{f}_{\sigma}$, where $\hat{b}$ and $\hat{f}_{\sigma}$ are usual
boson and fermion operators, respectively. 
This scheme introduces a gauge field, which requires a gauge fixing 
and the introduction of a Faddeev-Popov determinant \cite{leguillou95}. Gauge fluctuations 
should be taken into account beyond mean-field theory and the slave particles 
need to be convoluted to form the original fermionic operator $\tilde{c}$. 

An alternative approach is to employ the Faddeev-Jackiw \cite{faddeev88} and Dirac \cite{dirac50,sundermeyer82}  
theories for constrained systems and to develop a path integral representation 
for the Hubbard operators \cite{foussats00}. 
The partition function $Z$ is given in the Euclidean form as follows: 
\begin{eqnarray}
Z&=&\int {\cal D}X_{i}^{\alpha \beta}\;
\delta\left( X_{i}^{0 0} + \sum_{\sigma} X_{i}^{\sigma \sigma}-1 \right) \;  
\delta\left( X_{i}^{\sigma \sigma'} - \frac{X_{i}^{\sigma 0}
X_{i}^{0 \sigma'}}{X_{i}^{0 0}} \right) \nonumber \\
&\times&({\rm sdet} M_{AB})_{i}^{\frac{1}{2}} \exp\left(- \int d\tau\;L_E(X,
\dot{X})\right)\;, \label{part}
\end{eqnarray}
where fermionlike and bosonlike X-operators are described by 
Grassmann and usual complex variables, respectively,
$\tau=it$ and $\dot{X}=\partial_{\tau} X$. Note that we have removed the hat symbol  
of the Hubbard operators, because they become classical fields in the path integral formulation. 
The Euclidean Lagrangian $L_E(X,\dot{X})$ is given by 
\begin{eqnarray}
L_E(X, \dot{X}) =  \frac{1}{2}
\sum_{i, \sigma}\frac{(\dot{X_{i}}^{0 \sigma}\;X_{i}^{\sigma 0}
+ \dot{X_{i}}^{\sigma 0}\;X_{i}^{0 \sigma})}
 {X_{i}^{0 0}}
+ H(X)\;. 
\label{lagr}
\end{eqnarray}
The superdeterminant
\begin{equation}
({\rm sdet} M_{AB})_{i}^{\frac{1}{2}}=(-X^{00}_{i})^2 \,
\label{superdet}
\end{equation}
is equivalent to the determinant of the Dirac matrix \cite{dirac50,sundermeyer82} formed by the constraints 
and also to the Faddeev-Jackiw theory, where the symplectic matrix 
has even and odd Grassmann  elements (see Ref.~\onlinecite{foussats99} and references therein).  
The two constraints specified by the two $\delta$-functions in \eq{part} are necessary to 
recover the correct algebra of the original Hubbard operators \cite{foussats00}. 
The method used here recovers \cite{foussats99} the coherent state path-integral representation
for the $t$-$J$ model \cite{wiegmann88}. 

The partition function (\ref{part}) looks different from that usually found in other solid state problems. 
The measure of the integral contains the two constraints as well
as a superdeterminant $( {\rm sdet} M_{AB})_{i}^{\frac{1}{2}}$.  
In addition, the kinetic term of the Lagrangian (\ref{lagr}) is nonpolynomial. 
As a result, even if we discard the Heisenberg interaction and the long-range Coulomb interaction 
in the $t$-$J$ model, 
the theory is still highly nontrivial. In fact, the electron-electron interactions also come 
from the algebra of the X-operators in the $t$-$J$ model. 

One of the important aspects of the $t$-$J$ model is that charge carriers are absent 
at half-filling. To describe this explicitly, we rewrite 
the terms $X_{i}^{\sigma \sigma} X_{j}^{\sigma' \sigma'}$ in \eq{eq:H} 
in terms of $X_{i (j)}^{00}$ by using the local constraint 
$X_{i (j)}^{0 0} + \sum_{\sigma} X_{i (j)}^{\sigma \sigma}=1$. 
This local constraint is then kept by introducing the Lagrange multiplier $\lambda_{i}$: 
\be
\delta\left( X_{i}^{0 0} + \sum_{\sigma} X_{i}^{\sigma \sigma}-1 \right)=\int {\cal D} \lambda_{i} 
{\rm exp}\left(
{\mathrm i} \lambda_{i} \left( X_{i}^{0 0} + \sum_{\sigma} X_{i}^{\sigma \sigma}-1 \right) 
\right) \,.
\ee
The remaining two terms, 
$X_{i}^{\sigma \sigma'}X_{j}^{\sigma' \sigma}$ in the $J$ term and the term of the chemical potential, 
are replaced by integrating out the bosonic variable $X^{\sigma \sigma'}$ using the 
second $\delta$-function in \eq{part}, i.e., 
$X_{i}^{\sigma \sigma'} = (X_{i}^{\sigma 0}
X_{i}^{0 \sigma'})/X_{i}^{0 0}$. 
The resulting model is then described by 
$X_{i}^{\sigma 0}$, $X_{i}^{0 \sigma}$, $X_{i}^{0 0}$, and $\lambda_{i}$, and their mutual interactions. 

To achieve the systematic analysis of the $t$-$J$ model, we employ 
a nonperturbative technique based on a large-$N$ expansion \cite{foussats02},
where $N$ is the number of electronic degrees of freedom per site
and 1/$N$ is assumed to be a small parameter.  
We then extend the spin index $\sigma$ to a new index $p$ running from $1$ to $N$.
To get a finite theory in the limit of $N\rightarrow \infty$, 
we rescale the hopping $t_{ij}$ to $t_{ij}/N$, $J$ to $J/N$,
and $V_{i j}$ to $V_{i j}/N$. 
The completeness condition, i.e., the local constraint,  
becomes the $N$-extended one: $X_{i}^{0 0} + \sum_{p} X_{i}^{p p} = N/2 $. 
From this completeness condition, we 
can see that  $X^{00}$ is 
$O(N)$ and $X^{pp}$ is $O(1)$.   
As a consequence of this, the large-$N$ approach weakens the effective
spin interaction compared with the one associated with the charge degrees of freedom. 
In this sense, the large-$N$ scheme is potentially interesting in light of 
the new x-rays experiments \cite{ishii05,ghiringhelli12,chang12,achkar12,wslee14,ishii14,da-silva-neto15,da-silva-neto16,dellea17, ishii17,hepting18,jlin20,singh20,nag20}, 
which show unexpectedly strong excitations in the pure charge-channel. 

We write the boson fields $X_{i}^{00}$ and $\lambda_i$ in terms of
static mean-field values, $(r_0, \lambda_0)$, and dynamic fluctuations, $(\delta R_{i}, \delta\lambda_{i})$: 
\begin{eqnarray}
&&X_{i}^{0 0} = N r_{0}(1 + \delta R_{i}) \,\label{hubbop},\\
&&\lambda_{i} = \lambda_{0}+ \delta{\lambda_{i}} \, .
\label{hubbop-lambda}
\end{eqnarray}
For the fermionlike fields we write as follows: 
\begin{eqnarray}
f^{+}_{i p} &=& \frac{1}{\sqrt{N r_{0}}}X_{i}^{p 0} \,,
\\
f_{i p} &=& \frac{1}{\sqrt{N r_{0}}}\;X_{i}^{0 p} \,.
\label{fermoper}
\end{eqnarray}
Using \eq{hubbop} and the completeness condition, we obtain $r_0=\delta/2$,  
where $\delta$ is the hole doping rate away from half-filling. 
The fermion variable $f_{ip}$ is proportional to $X_{i}^{0 p}$ 
and should not be associated with the so-called spinon  
in the slave-boson method.

$X^{0 0}$ to the second power appears in \eq{superdet} because we were working with two spin 
projections. 
After the extension to large-$N$,  
the superdeterminant becomes   
\begin{eqnarray}
({\rm sdet} M_{AB})_{i}^{\frac{1}{2}} = {(Nr_{0})}^N \left[-
(1 + \delta R_i)\right]^{N}\;.  
\label{supdetext}
\end{eqnarray}
The constant factor $(Nr_{0})^N$ contributes to the path-integral normalization factor. 
On the other hand, the term $[-(1 + \delta R_i)]^{N}$ can be written in terms of $N$ complex bosonic  
ghost fields ${\cal Z}_{i p}$ in an integral representation (Ref.~\onlinecite{foussats02}). 
Therefore the superdeterminant contributes to the effective
Lagrangian as 
\begin{eqnarray}
L_{ghost}({\bf {\cal Z}}) = - \sum_{i p}\;
{\bf {\cal Z}}^{\dag}_{i p}\left(\frac{1}{1 + \delta R_i}\right)
{\bf {\cal Z}}_{i p}\;.
\end{eqnarray}

The  exchange  interaction term ($J$ term) contains the four fermion fields and the two bosons in 
the denominator. They can be decoupled 
through a Hubbard-Stratonovich transformation 
by introducing the field $\Delta_{ij}$,  
\be
\Delta_{ij} = \frac{J}{4N} \sum_{p} \frac{f^{\dag}_{j p} f_{i p}}{ \sqrt{(1 + \delta R_{i})
(1 + \delta R_{j})}} \,. 
\label{bond-field}
\ee
This field describes bond-charge fluctuations because 
the fermions with the same spin projection $p$ sit on the nearest-neighbor sites and 
the sum over $p$ is taken. This bond field $\Delta_{ij}$ is parameterized as 
\be
\Delta_i^{\eta}=\Delta(1+r_i^\eta+iA_i^\eta)\,,
\label{staticDelta}
\ee
where $r_i^{\eta}$ and $A_i^{\eta}$ correspond to the real and imaginary parts of 
the bond-field fluctuations, respectively, and $\Delta$ is a static mean-field value. 
The index $\eta$ denotes the bond directions
${\eta}_{1}=(1,0)$ and ${\eta}_{2}=(0,1)$ on a square lattice. 
Finally, we expand the term $1/(1+\delta R)$ in powers of $\delta R$, which generates various interactions  
between fermions and bosons. The number of the interactions considered in theory is controlled in powers 
of $1/N$. 

We may define a six-component boson field 
\be
\delta X^{a} = (\delta
R\;,\;\delta{\lambda},\; r^{{\eta}_{1}},\;r^{{\eta}_{2}}
,\; A^{{\eta}_{1}},\;
A^{{\eta}_{2}})\,,
\label{deltaXa}
\ee
where the  first component $\delta R$ 
is related to on-site charge fluctuations and $\delta \lambda$ is the fluctuation of the
Lagrange multiplier $\lambda_i$ associated
with the completeness condition; the remaining components come from fluctuations 
of the bond field [see \eq{staticDelta}]. 
Hence the effective theory can be described in terms of  boson fields $\delta X^{a}$, 
fermion fields $f_{i p}$, ghost fields ${\bf {\cal Z}}_{i p}$, and their mutual interactions. 

\begin{figure}
\vspace{1cm}
\begin{center}
\setlength{\unitlength}{1cm}
\includegraphics[width=8.5cm]{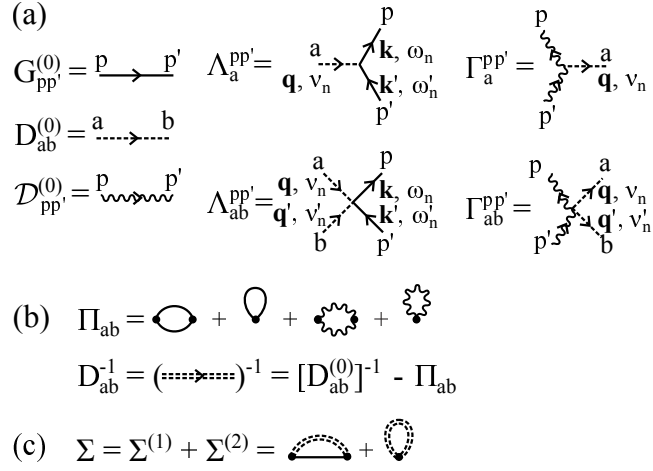}
\end{center}
\caption{(a) Propagators and vertices in a large-$N$ theory up to $O(1/N)$. Solid line represents 
the fermion propagator $G^{(0)}_{pp'}$; dashed line the $6 \times 6$ boson propagator
$D^{(0)}_{ab}$ for the six-component fields $\delta X^a$; wavy line the ghost propagator 
${\cal D}^{(0)}_{pp'}$. 
$\Lambda^{pp'}_a$ and  $\Gamma_{a}^{pp'}$ describe the three-point interaction
between two fermions with spin indices $p$ and $p'$ and one boson with the component $a$, and 
two ghost fields with one boson, respectively. 
$\Lambda^{pp'}_{ab}$ and ${\Gamma}_{ab}^{pp'}$ are the four-point interaction among the two fermions and two bosons,  
and two ghost fields and two bosons, respectively. 
(b) The irreducible boson self-energy $\Pi_{ab}$ and the renormalized boson propagator $D_{ab}$. 
(c) The electron self-energy $\Sigma({\bf k},\omega)$ at order of $1/N$. 
} 
\label{FD}
\end{figure}

From the quadratic part for fermions we obtain the bare electron 
propagator in the paramagnetic phase (solid line in \fig{FD}), 
\begin{eqnarray}
G^{(0)}_{pp} ({\bf k}, {\rm i}\omega_{n}) = \frac{1}{{\rm i}\omega_{n} -\varepsilon_{\vk}}\,.
\label{G0-A}
\end{eqnarray}
Here ${\vk}$ and ${\rm i}\omega_{n}$ are momentum and fermionic  Matsubara frequency, respectively.
The bare fermionic propagator $G^{(0)}$ is $O(1)$. The electron dispersion $\varepsilon_{\vk}$ is 
written as 
\be
\varepsilon_{\vk} = \varepsilon_{\vk}^{\parallel}  + \varepsilon_{\vk}^{\perp} \,,
\label{xik-A}
\ee
where the in-plane dispersion $\varepsilon_{\vk}^{\parallel}$ and the out-of-plane dispersion 
$\varepsilon_{\vk}^{\perp}$ are given, respectively, by
\be
\varepsilon_{\vk}^{\parallel} = -2 \left( t \frac{\delta}{2}+\Delta \right) (\cos k_{x}+\cos k_{y})-
4t' \frac{\delta}{2} \cos k_{x} \cos k_{y} - \mu \,,\\
\label{Epara-A}
\ee
\be
\varepsilon_{\vk}^{\perp} = - 2 t_{z} \frac{\delta}{2} (\cos k_x-\cos k_y)^2 \cos k_{z}  \,.
\label{Eperp-A}
\ee
Here $\lambda_0$ in \eq{hubbop-lambda} was absorbed in the chemical potential $\mu$. 
In-plane momenta $k_x$ and $k_y$ and out-of-plane momentum $k_z$ are measured 
in units of $a^{-1}$ and $d^{-1}$, respectively; $a$ is the lattice constant in the square lattice and 
$d$ is the distance between the planes. 
The quantity $\Delta$ is given  by the expression 
\bea{\label {Delta-A}}
\Delta = \frac{J}{4N_s N_z} \sum_{\vk} (\cos k_x + \cos k_y) n_F(\varepsilon_\vk) \; , 
\eea
where $n_F$ is the Fermi function, $N_s$ is the total 
number of lattice sites on the square lattice, and $N_z$ is the number of layers along the $z$ direction. 
For a given doping $\delta$, $\mu$, and $\Delta$ are determined self-consistently by solving \eq{Delta-A} and 
\be
(1-\delta)=\frac{2}{N_s N_z} \sum_{\vk} n_F(\varepsilon_\vk)\,.
\ee

From the quadratic part of the six-component boson field $\delta X^{a}$,
we obtain the  inverse of the bare bosonic propagator (dashed line in \fig{FD}),
connecting two components $a$ and $b$,
\begin{widetext}
\begin{eqnarray}
\left[D^{(0)}_{ab}({\bf q},\mathrm{i}\nu_{n})\right]^{-1}= N \left(
 \begin{array}{cccccc}
\frac{\delta^2}{2} \left[ V(\vq)-J(\vq)\right]
& r_{0} & 0 & 0 & 0 & 0 \\
   r_{0} & 0 & 0 & 0 & 0 & 0 \\
   0 & 0 & \frac{4}{J}\Delta^{2} & 0 & 0 & 0 \\
   0 & 0 & 0 & \frac{4}{J}\Delta^{2} & 0 & 0 \\
   0 & 0 & 0 & 0 & \frac{4}{J}\Delta^{2} & 0 \\
   0 & 0 & 0 & 0 & 0 & \frac{4}{J}\Delta^{2} \
 \end{array}
\right)  \,,
\label{D0-A}
\end{eqnarray}
\end{widetext}
where $D^{(0)}_{ab}$ is $O(1/N)$, 
$\vq$ a three-dimensional wavevector, $\nu_n$ a bosonic Matsubara frequency, 
and $J(\vq) = \frac{J}{2} (\cos q_x +  \cos q_y)$. 
$V(\vq)$ is the long-range Coulomb interaction in momentum space for a layered system 
\be
V(\vq)=\frac{V_c}{A(q_x,q_y) - \cos q_z} \,,
\label{LRC-A}
\ee
where $V_c= e^2 d(2 \epsilon_{\perp} a^2)^{-1}$ and 
\be
A(q_x,q_y)=\alpha (2 - \cos q_x - \cos q_y)+1 \,.
\ee
These expressions are easily obtained by solving Poisson's equation on the lattice \cite{becca96}.  
Here $\alpha=\frac{\tilde{\epsilon}}{(a/d)^2}$, $\tilde{\epsilon}=\epsilon_\parallel/\epsilon_\perp$,  
and $\epsilon_\parallel$ and $\epsilon_\perp$ are the 
dielectric constants parallel and perpendicular to the planes, respectively; 
$e$ is the electric charge of electrons.  

From the quadratic part of the ghost fields ${\bf {\cal Z}}_{i p}$, 
we obtain the  inverse of the bare ghost propagator (wavy line in Fig.\ref{FD}) 
\begin{eqnarray}
{\cal D}^{(0)}_{pp'} = - \delta_{pp'}, 
\end{eqnarray}
which is $O(1)$.

The interactions between $\delta X^a$, $f_{p}$, and  ${\bf {\cal Z}}_{p}$ are given by three- and four-point vertices. 
The three-point vertices read  
\begin{eqnarray}\label{gamm3}
&&\Lambda^{pp'}_{a} =  (-1) \left(\frac{i}{2}(\omega_n +
{\omega'}_n) + \mu + 2\Delta \left[
\cos \left( k_x -\frac{q_x}{2} \right) \cos\frac{q_x}{2} + \cos \left( k_y -\frac{q_y}{2} \right) \cos\frac{q_y}{2} \right]  ; 
\;1; \right. \nonumber \\
&&
 \left.  
 - 2\Delta \cos \left( k_x-\frac{q_x}{2} \right) \; ;
- 2\Delta \cos \left(k_y-\frac{q_y}{2} \right) ; \;  
 2 \Delta \sin \left(k_x-\frac{q_x}{2} \right) ;\; 
 2 \Delta \sin \left(k_y-\frac{q_y}{2} \right) \right)  \delta^{pp'},  \nonumber \\
 \end{eqnarray}
and 
\begin{eqnarray}
\Gamma_{a}^{pp'} = (-1) \delta_{a1}\delta_{pp'}\, .
\end{eqnarray}
They represent the interaction between two fermions and one boson, and two ghosts and one boson,
respectively. 
The four-point vertices, $\Lambda^{pp'}_{ab}$ and  $\Gamma_{ab}^{pp'}$, describe the interaction between
two fermions and two bosons, and two bosons and two ghosts, respectively. 
$\Lambda^{pp'}_{ab}$ fulfills the symmetry of 
$\Lambda^{pp'}_{ab} = \Lambda^{pp'}_{ba}$. 
The elements different from zero are:
\begin{eqnarray}
&&\Lambda^{pp'}_{\delta R \delta R}=  \left(\frac{i}{2} (\omega_n + {\omega'}_n)
+ \mu  \right. \nonumber \\
&& \hspace{10mm} \left. + \Delta \sum_{\eta=x,y}
\cos \left( k_\eta-\frac{q_\eta+q'_\eta}{2} \right)\;
\left( \cos\frac{q_\eta}{2} \; \cos\frac{q'_\eta}{2}\;
+\; \cos\frac{q_\eta+q'_\eta}{2} \right) \right)\delta^{pp'},
\label{Labpp'}
\end{eqnarray}
\begin{eqnarray}
\Lambda^{pp'}_{\delta R \delta\lambda}=\frac{1}{2}
\;\delta^{pp'},
\label{Labpp'2}
\end{eqnarray}
\begin{eqnarray}
\Lambda^{pp'}_{\delta R \; r^{\eta}}= -\Delta 
\cos \left( k_\eta-\frac{q_\eta+q'_\eta}{2} \right) \cos\frac{q'_\eta}{2} \; 
\delta^{pp'},
\label{Labpp'3}
\end{eqnarray}
and
\begin{eqnarray}
\Lambda^{pp'}_{\delta R \; A ^{\eta}}= \Delta 
\sin \left( k_\eta-\frac{q_\eta+q'_\eta}{2} \right) \cos\frac{q'_\eta}{2} \; 
\delta^{pp'}.
\label{Labpp'4}
\end{eqnarray}
The four legs interaction vertex $\Gamma_{ab}^{pp'}$ reads
\be
{\Gamma}_{ab}^{pp'} = \delta_{a1} \delta_{b1} \delta_{p p'} \, .
\ee
Each vertex conserves the momentum and energy and 
it is at $O(1)$. 

The chemical potential $\mu$ appears in Eqs.~(\ref{gamm3}) and (\ref{Labpp'}). 
Its origin is easily traced. 
First, the $N$-extended chemical potential term $\mu \sum_{i,p} {X}_{i}^{pp}$ is 
written as $\mu \sum_{i,p}  (X_{i}^{p 0} X_{i}^{0 p})/X_{i}^{0 0}$ by using 
the second $\delta$-function in \eq{part}. 
Second, its denominator is expanded in powers of  
$\delta R_{i}$ [see also \eq{hubbop}]. 
Third, to compute quantities up to $O(1/N)$ 
the expansion is made up to $\delta R_{i}^2$, which generates a term $\mu$ in 
the three-legs and four-legs vertices. 

The propagators and vertices are summarized in \fig{FD}(a). 
For a large-$N$ approach, any  physical
quantity can be calculated at a given order by counting the
powers of $1/N$ in vertices and propagators involved in the
corresponding Feynman diagram. In \fig{FD}(a), we show the 
necessary diagrams to  calculate quantities up to $O(1/N)$. 
When going beyond that order, we need to collect more vertices in the expansion of 
the expression $1/(1+\delta R)$, 
which demonstrates one of nontrivial aspects of the $t$-$J$ model. 

So far, all procedures are exact formally and no approximations are introduced. 
Now we show calculations up to the order of $1/N$.  
The bare susceptibility $D^{(0)}_{ab}$ is already at order of $1/N$. 
From the Dyson equation,  the bosonic propagator (dashed line 
in Fig.\ref{FD}) is renormalized at the same order as 
\be
[D_{ab}(\vq,\mathrm{i}\nu_n)]^{-1} 
= [D^{(0)}_{ab}(\vq,\mathrm{i}\nu_n)]^{-1} - \Pi_{ab}(\vq,\mathrm{i}\nu_n)\,,
\label{dyson-A}
\ee
where the $6 \times 6$ boson self-energy
matrix $\Pi_{ab}$ is evaluated by the diagrams shown in Fig. \ref{FD}(b): 
\begin{eqnarray}
&& \Pi_{ab}(\vq,\mathrm{i}\nu_n)
            = -\frac{N}{N_s N_z}\sum_{\vk} h_a(\vk,\vq,\varepsilon_\vk-\varepsilon_{\vk-\vq}) 
            \frac{n_F(\varepsilon_{\vk-\vq})-n_F(\varepsilon_\vk)}
                                  {\mathrm{i}\nu_n-\varepsilon_\vk+\varepsilon_{\vk-\vq}} 
            h_b(\vk,\vq,\varepsilon_\vk-\varepsilon_{\vk-\vq}) \nonumber \\
&& \hspace{25mm} - \delta_{a\,1} \delta_{b\,1} \frac{N}{N_s N_z}
                                       \sum_\vk \frac{\varepsilon_\vk-\varepsilon_{\vk-\vq}}{2}n_F(\varepsilon_\vk) \; .
                                       \label{Pi-A}
\end{eqnarray}
The vertices $h_a$ are given by 
\begin{align}
 h_a(\vk,\vq,\nu) =& \left\{
                   \frac{2\varepsilon_{\vk-\vq}+\nu+2\mu}{2}+
                   2\Delta \left[ \cos\left(k_x-\frac{q_x}{2}\right)\cos\left(\frac{q_x}{2}\right) +
                                  \cos\left(k_y-\frac{q_y}{2}\right)\cos\left(\frac{q_y}{2}\right) \right];
                 \right. \nonumber \\
               & \hspace{-10mm} \left.1;  -2\Delta \cos\left(k_x-\frac{q_x}{2}\right); -2\Delta \cos\left(k_y-\frac{q_y}{2}\right);
                         2\Delta \sin\left(k_x-\frac{q_x}{2}\right);  2\Delta \sin\left(k_y-\frac{q_y}{2}\right)
                 \right\} \, .
\label{vertex-h-A}
\end{align}
Here the dependence on $k_z$ and $q_z$ enters only through $\epsilon_{\vk-\vq}$  in the first column 
in \eq{vertex-h-A}, whereas the other columns contain the in-plane momentum only. 
It is important to note that the  only role played by the ghost fields
is to cancel infinities coming from  the frequency dependence of the vertices involved in 
the first two diagrams in Fig. \ref{FD}(b).

The Green's function acquires dynamical corrections as shown in \fig{FD}(c). 
The self-energy consists of two parts at order of $1/N$: 
\begin{equation}
\Sigma({\mathbf{k}},\mathrm{i}\omega_{n})=
\Sigma^{(1)}({\mathbf{k}},\mathrm{i}\omega_{n})+
\Sigma^{(2)}({\mathbf{k}},\mathrm{i}\omega_{n}) \; ,
\label{Sigma-A}
\end{equation}
where 
\begin{eqnarray}\label{Sigma1-A}
\Sigma^{(1)}({\mathbf{k}},\mathrm{i}\omega_{n})&=&\frac{1}{N_{s}N_{z}}\sum_{{\mathbf{q}}
,\nu_{n}} \sum_{ab} \Lambda^{pp}_{a}
\;D_{ab}({\mathbf{q}},\mathrm{i}\nu_{n}) \nonumber\\
&\times& G_{pp}^{(0)} ({\mathbf{k-q}},\mathrm{i} \omega_{n}- \mathrm{i} \nu_{n}) \;
\Lambda^{p p}_{b} \,,
\end{eqnarray}
and 
\begin{eqnarray}\label{Sigma2-A}
\Sigma^{(2)}({\mathbf{k}},\mathrm{i}\omega_{n})&=&\frac{1}{N_{s}N_z}\sum_{{\mathbf{q}}
,\nu_{n}}  \sum_{ab} \Lambda^{pp}_{ba}\;
D_{ab}({\mathbf{q}},\mathrm{i}\nu_{n}).
\end{eqnarray}
In Eqs.~(\ref{Sigma-A})-(\ref{Sigma2-A}) we have omitted the index $p$ in the self-energy for simplicity.  
$\Sigma^{(1)}(\vk,\mathrm{i}\omega_n)$ corresponds to the Fock diagram, the first one in \fig{FD}(c), 
and $\Sigma^{(2)}(\vk,\mathrm{i}\omega_n)$ the Hartree diagram, 
the second one in \fig{FD}(c). Usually the Hartree term gives 
a constant contribution, which can be absorbed in the chemical potential, but here 
both diagrams must be considered because of the frequency and momentum 
dependence of the vertices $\Lambda^{pp}_{a}$ and $\Lambda^{pp}_{ab}$.   
One can easily see in Eqs.~(\ref{Sigma1-A}) and (\ref{Sigma2-A}) that 
the obtained self-energy is indeed at order of $1/N$: 
$G_{pp}^{(0)}$ and the vertices are at $O(1)$, and $D_{ab}$ is at $O(1/N)$.

Using the spectral representation of $D_{ab}$, we obtain
\begin{eqnarray}\label{Sigma11}
\Sigma^{(1)}({\mathbf{k}},\mathrm{i}\omega_{n})&=&-\frac{1}{\pi N_{s}N_z}\int
d\nu \sum_{{\mathbf{q}}
,\nu_{n}} \sum_{ab}  \Lambda_{a}^{pp}
\frac{{\rm Im}D_{ab}(\vq,\nu)}{\mathrm{i}\nu_{n}-\nu}\;
\Lambda_{b}^{pp} \nonumber \\
&\times& G_{pp}^{(0)}({\mathbf{k-q}},\mathrm{i}\omega_{n} - \mathrm{i}\nu_{n}), 
\end{eqnarray}
\begin{equation}\label{Sigma21}
\Sigma^{(2)}({\mathbf{k}},\mathrm{i}\omega_{n})=-\frac{1}{\pi N_{s}N_z}\int
d\nu \sum_{{\mathbf{q}} ,\nu_{n}} \sum_{ab} 
\Lambda_{ba}^{pp} \;\frac{{\rm Im} D_{ab}(\vq,\nu)} 
{\mathrm{i}\nu_{n}-\nu} \, .
\end{equation}
After performing the Matsubara sum and the 
analytical continuation, 
the imaginary part of $\Sigma$ can be written in a compact form, 
\begin{eqnarray}\label{SigmaIm}
    {\mathrm{Im}}\Sigma({\mathbf{k}},\omega) &=& -\frac{1}{N_{s}N_z}
\sum_{{\mathbf{q}}} \sum_{ab}  h_{a}(\vk,\vq,\omega-\varepsilon_{{\vk-\vq}}) \nonumber\\
&& \times {\rm Im} D_{ab}(\vq,\omega-\varepsilon_{{\vk-\vq}}) \; h_{b}(\vk,\vq,\omega-\varepsilon_{{\vk-\vq}}) \nonumber\\
&&\times [n_{F}( -\varepsilon_{\vk-\vq}) +n_{B}(\omega-\varepsilon_{{\vk-\vq}})] \,,
\end{eqnarray}
where $n_B$ is the Bose factor.   
We obtain $\mathrm{Re} \Sigma({{\bf{k}}},\omega)$ from  $\mathrm{Im} \Sigma({{\bf{k}}},\omega)$  via 
the Kramers-Kronig relations. 
Since the electron Green's function $G(\vk ,\omega)$ is given by 
\be
G(\vk, \omega)^{-1} = \omega +\mathrm{i} \eta - \varepsilon_{\vk} - \Sigma (\vk, \omega) \, ,
\ee
with $\eta (>0)$ being infinitesimally small, the spectral function $A({\bf k},\omega)=-\frac{1}{\pi} {\rm Im}G({\bf k},\omega)$ is obtained as 
\begin{eqnarray}\label{A}
A({\bf k},\omega)= -\frac{1}{\pi}\frac{{\rm Im}\Sigma({\bf k},\omega)-\eta}
{[\omega- \varepsilon_{{\bf k}}-{\rm Re}\Sigma({\bf k},\omega)]^2
+ [-{\rm Im}\Sigma({\bf k},\omega) +\eta]^2} \,.
\label{ak}
\end{eqnarray}

\section{Additional results after particle-hole transformation} 
\begin{figure}[th]
\centering
\includegraphics[width=8.5cm]{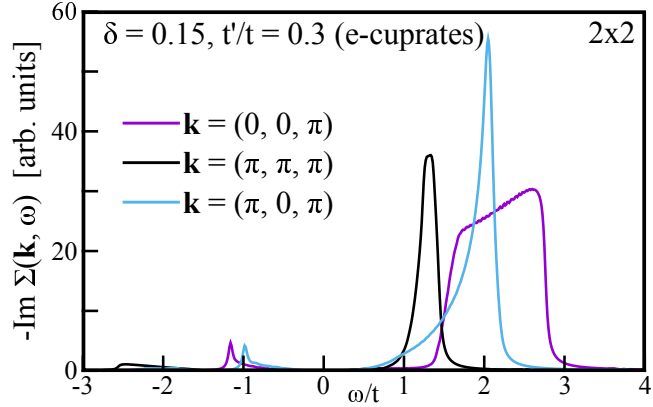}
\caption{(Color online) 
Imaginary part of the electron self-energy, $-{\rm Im}\Sigma(\vk,\omega)$, as a function of $\omega$ 
for several choices of $\vk$ after the particle-hole transformation of \fig{ImSig-asym}. 
}
\label{ImSig-asym-ph}
\end{figure}

\begin{figure}[ht]
\centering
\includegraphics[width=8cm]{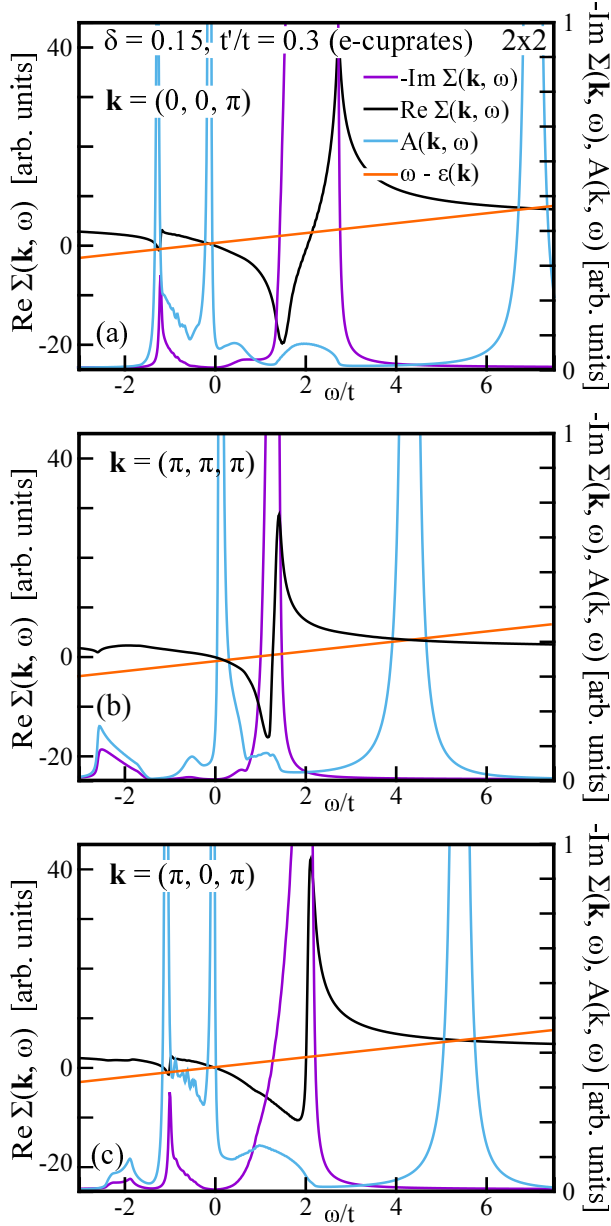}
\caption{(Color online) 
Energy dependence of $- {\rm Im}\Sigma(\vk,\omega)$, ${\rm Re}\Sigma(\vk,\omega)$, 
and $A(\vk,\omega)$ at $\vk=(0,0,\pi)$ (a), $(\pi,\pi,\pi)$ (b), and  $(\pi, 0,\pi)$ (c) 
after the particle-hole transformation of \fig{ImReA}. 
}
\label{ImReA-ph}
\end{figure}

In Sec.~III F, we have presented results after the particle-hole transformation so that 
they will be compared directly with experiments for e-cuprates. 
It is also informative to present the $\omega$ dependence of the self-energy 
and the spectral function. 
Hence we perform the particle-hole transformation of Figs.~\ref{ImSig-asym} and \ref{ImReA} 
and present here the corresponding results in Figs.~\ref{ImSig-asym-ph} and \ref{ImReA-ph}, respectively. 

\begin{figure}[ht]
\centering
\includegraphics[width=9cm]{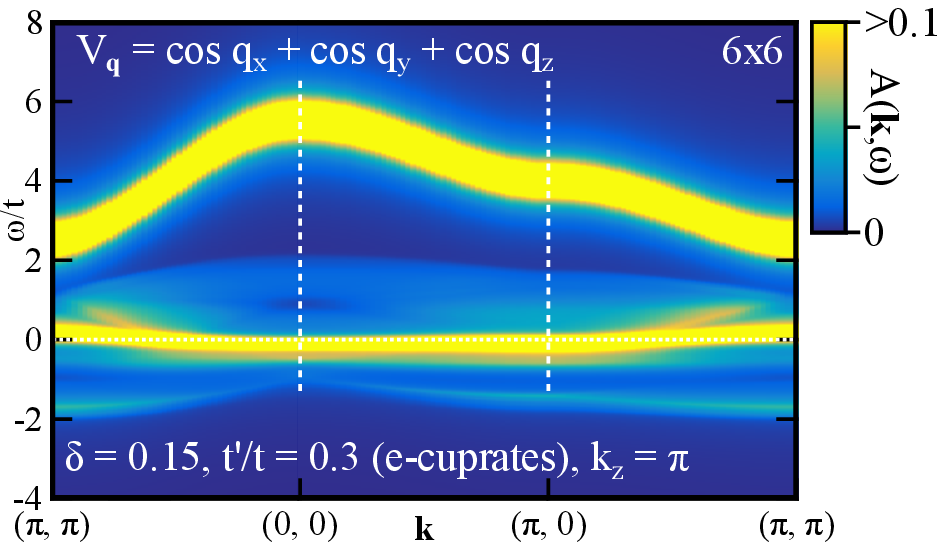}
\caption{(Color online) 
Intensity map of $A(\vk,\omega)$ in the absence of the long-range Coulomb interaction 
along the direction $(\pi,\pi)$-$(0,0)$-$(\pi,0)$-$(\pi,\pi)$ 
in the {\it particle} picture; 
both on-site charge and bond-charge fluctuations are taken into account. 
$k_z$ dependence is weak and $k_z=\pi$ is taken as a representative value. 
The corresponding result in the {\it hole} picture is presented in \fig{Akw-map-SR}. 
}
\label{Akw-map-SR-ph}
\end{figure}

We also present the corresponding result to \fig{Akw-map-SR} 
in \fig{Akw-map-SR-ph}. This result together with \fig{Akw-ph}(c) will be useful 
when one wishes to discuss the role of the long-range Coulomb interaction 
on the basis of the one-particle excitation spectrum for e-cuprates.

\newpage

\bibliography{main} 

\begin{thebibliography}{104}
\bibitem [{mis({\natexlab{a}})}]{misc-ph}%
  \BibitemOpen
  \href@noop {} {} \bibinfo {note} {{It is possible to
  define a particle-hole transformation with $\vQ=(\pi,\pi,\pi)$. Even in this
  case, our conclusions are not modified.}}\BibitemShut {Stop}%
\bibitem [{mis({\natexlab{b}})}]{misc-factor2}%
  \BibitemOpen
  \href@noop {} {} \bibinfo {note} {A factor of $1/2$ here
  comes from a large-$N$ formalism where $t$ is scaled by $1/N$. We may assume
  $N=2$ in comparison with experiments.}\BibitemShut {Stop}%
\end{thebibliography}

\end{document}